\documentclass[fleqn,usenatbib]{mnras}

\usepackage{newtxtext,newtxmath}

\usepackage[T1]{fontenc}

\DeclareRobustCommand{\VAN}[3]{#2}
\let\VANthebibliography\thebibliography
\def\thebibliography{\DeclareRobustCommand{\VAN}[3]{##3}\VANthebibliography}

\usepackage{graphicx}	
\usepackage{amsmath}	
\usepackage[utf8]{inputenc}
\usepackage{caption}
\usepackage{subcaption}
\usepackage{xspace}
\usepackage{tikz}
\usepackage{ulem}
\usepackage{comment}
\usepackage{xcolor}

\newcommand{\cm}{\,\ifmmode{{\mathrm{cm}}}\else cm\fi}

\newcommand{\ergps}{\,{\rm erg}\,{\rm s}\ifmmode{}^{-1}\else${}^{-1}$\fi}
\newcommand{\Mpch}{\,{\rm Mpc}\,\ifmmode h^{-1}\else $h^{-1}$\fi}
\newcommand{\snru}{\,\ifmmode{\mathrm{Myr}^{-1}}\else Myr${}^{-1}$\fi}
\newcommand{\kms}{\,\ifmmode{\mathrm{km}\,\mathrm{s}^{-1}}\else km\,s${}^{-1}$\fi\xspace}
\newcommand{\teddy}{\,\ifmmode{{t_\mathrm{eddy}}} \else $t_{\rm eddy}$\fi\xspace}
\newcommand{\tcoolmix}{\,\ifmmode{{t_\mathrm{cool, mix}}} \else $t_{\rm cool, mix}$\fi\xspace}
\newcommand{\tcoolcold}{\,\ifmmode{{t_\mathrm{cool, cold}}} \else $t_{\rm cool, cold}$\fi\xspace}

\newcommand{\Da}{\ifmmode{\mathrm{Da}}\else Da\fi \xspace}
\newcommand{\Ma}{\ifmmode{\mathcal{M}_{\mathrm{A}}}\else $\mathcal{M}_{\mathrm{A}}$\fi \xspace}
\newcommand{\Ms}{\ifmmode{\mathcal{M}_{\mathrm{s}}}\else $\mathcal{M}_{\mathrm{s}}$\fi \xspace}
\newcommand{\dd}{\mathrm{d}}

\newcommand{\textchange}[1]{#1}

\title[Multiphase turbulence and Magnetic fields]{Magnetic Fields in Multiphase Turbulence: Impacts on Dynamics and Structure}

\author[Das \& Gronke]{
Hitesh Kishore Das,$^{1}$\thanks{E-mail: hitesh@mpa-garching.mpg.de}
Max Gronke,$^{1}$
\\
$^{1}$Max Planck Institute for Astrophysics, Garching 85748, Germany\\
}

\date{Draft from \today}

\pubyear{2023}

\begin{document}
\label{firstpage}
\pagerange{\pageref{firstpage}--\pageref{lastpage}}
\maketitle

\begin{abstract}
Both multiphase gas and magnetic fields are ubiquitous in astrophysics. 
However, the influence of magnetic fields on mixing of the different phases is still largely unexplored. In this study, we use both turbulent radiative mixing layer (TRML) and turbulent box simulations to examine the effects of magnetic fields on cold gas growth rates, survival, and the morphology of the multiphase gas. 
Our findings indicate that, in general, magnetic fields suppress mixing in TRMLs while turbulent box simulations show comparatively marginal differences in growth rates and survival of the cold gas.
We reconcile these two seemingly contrasting results by demonstrating that similar turbulent properties result in comparable mixing  -- regardless of the presence or absence of magnetic fields.
We, furthermore, find the cold gas clump size distribution to be independent of the magnetic fields but the clumps are more filamentary in the MHD case.
Synthetic MgII absorption lines support this picture being marginally different with and without magnetic fields; both cases aligning well with observations. We also examine the magnetic field strength and structure in turbulent boxes. We generally observe a higher mean magnetic field in the cold gas phase due to flux freezing and reveal fractal-like magnetic field lines in a turbulent environment.
\end{abstract}




\section{Introduction}

The fact that most of the astrophysical media are multiphase in nature is a well-established one, from observational \citep[e.g.][]{Tumlinson2017TheMedium, Veilleux2020CoolImplications}, numerical, and theoretical investigations \citep{McKee1977ASubstrate., Donahue2022, Faucher-Giguere2023KeyMedium}. The multiphase nature of the interstellar (ISM), circumgalactic (CGM) and intracluster (ICM) medium is also expected to play an important role in the overall evolution of the associated systems, from the general baryon cycle to feedback processes \citep{Veilleux2005GalacticWinds, Peroux2020}. However, there are many aspects of multiphase media, \textchange{like survival and characteristic size of cold media}, that are \textchange{still} in question and are an active field of research.

Many forays towards understanding the multiphase gas use an idealised version of the medium. There are studies which focus on the development of the multiphase gas by condensation from an initially static hot ambient medium via thermal instability in 1D \citep{Sharma2012ThermalGalaxies, Waters2019Non-isobaricInstability}, 2D \citep{McCourt2018AGas} and 3D simulations \citep{Gronke2020IsMisty}. Such studies are a good way to isolate and investigate the role of different factors like magnetic fields \citep{Sharma2010ThermalClusters, 10.1093_mnras_sty293}, metallicity \citep{Das2021ShatterInstability}, gravity \citep{Mccourt2012ThermalHaloes}, density fluctuations \citep{Choudhury2019MultiphaseFluctuations}, rotation \citep{Sobacchi2019TheAnalysis} or cosmic rays \citep{Butsky2020TheMedium}.

It is also well known that the astrophysical media are highly turbulent, due to their high Reynolds number. This has been shown by many observational \citep{Elmegreen2004InterstellarProcesses, FalcetaGoncalves2014, VidalGarcia2021,Li2022TurbulenceGalaxy} and numerical studies \citep{Brandenburg2011AstrophysicalModeling, Federrath2013OnTurbulence, Burkhart2020TheCATS}. Hence, many studies like \citet{Mohapatra2019TurbulenceThermodynamics, Gronke2022SurvivalMedium, Mohapatra2022CharacterizingSimulations, Mohapatra2022MultiphaseDriving} investigate the evolution of the multiphase gas in the presence of a turbulent astrophysical media. Turbulence both amplifies and destroys multiphase gas. The density and temperature perturbations in a turbulent medium can enhance the creation of multiphase gas, while the same turbulent motions can mix the existing multiphase gas, which might further cool or mix away depending on the cooling timescale.

Not just astrophysics, multiphase turbulence is also a very relevant topic at more terrestrial scales and is also an active field of research in general fluid dynamics circles, as there are many applications like combustion dynamics, smoke transport and meteorology where multiphase interactions play a crucial role. One seminal result in the field was by \citet{Damkohler1940}, where they found that the behaviour of a flame front in a turbulent medium differs depending on the ratio of the reaction and turbulent timescales, i.e. the Damk\"ohler number. \citet{Tan2021RadiativeCombustion} further explore this parallel in the context of hydrodynamic turbulent radiative mixing layers (TRMLs) in astrophysical media.

Generally, there are three stages in the evolution of a turbulent multiphase medium, with many studies examining each. First, the production or presence of seed multiphase gas. The exact mechanism of this can vary from medium to medium. For the CGM, this can either be in the form of multiphase ISM transported into the CGM by feedback mechanism, or via condensations from the hot medium due to thermal instability. The second stage is the growth of one of the phases in the multiphase gas. And, the final stage is the equilibrium or steady state.

In order to understand the second stage of the multiphase gas evolution, \citet{Gronke2022SurvivalMedium} study the growth of cold gas in a thermally stable, ambient turbulent medium. They found a critical radius for the size of the seed cold gas cloud in a given turbulent hot ambient medium \citep[akin to the survival criterion previously found for laminar flows][]{Gronke2018TheWind, Li2020SimulationBackground, Kanjilal2021GrowthCooling}. Their results also agreed with the expectations from the previous hydrodynamics TRML results from \citet{Tan2021RadiativeCombustion}, indicating TRMLs might be the principle mechanism for mixing in a multiphase medium.

But, apart from being multiphase and turbulent, the astrophysical media are also known to possess substantial magnetic fields as seed primordial magnetic fields are amplified due to structure formation and other baryonic dynamics \citep{Dimopoulos1997EvolutionFields, Subramanian2015TheFields}. There are many observational evidences for ubiquitousness of magnetic fields using techniques like Faraday rotation \citep{Dreher1987TheGas, Kim1990TheGalaxies, Taylor1993MagneticCluster, Clarke2001AFields}, dust alignment \citep{Ade2015iPlanck/iDust}, and others \citep{LopezRodriguez2021}. And, many numerical studies also point to a similar prevalence of magnetic fields \citep[e.g.,][]{Pakmor2020MagnetizingGalaxies, vandeVoort2021}.

The presence of magnetic fields can be disruptive to mixing via TRMLs. It is well-known that linearly, the Kelvin-Helmholtz instability is suppressed for specific magnetic field orientation \citep{Chandrasekhar1961HydrodynamicStability}\textchange{, while} \citet{Ji2019SimulationsLayers} show that the nonlinear evolution of the instability with radiative cooling is suppressed for all orientations of the initial magnetic field. \footnote{This study was within the slow-cooling regime from \citet{Tan2021RadiativeCombustion}.} Hence, the inclusion of magnetic fields may change the overall evolution of multiphase gas, resulting in different survival criteria and cold gas growth rates.
In summary, while it has been shown in recent work that mixing and subsequent cooling can lead to the survival and even the production of cold gas, and thus explain the ubiquitous presence of multiphase gas in turbulent systems -- where this cold gas should be destroyed rapidly -- magnetic fields might ruin this picture by preventing mixing and hence hindering cooling.

In this paper, we investigate the influence of magnetic fields on the general phenomenon of mixing between the phases in a multiphase gas. For that purpose, we use two kinds of simulations, TRMLs and turbulent boxes, with and without magnetic fields. First, we expand on the parameter space for TRMLs explored in previous studies, to confirm the suppression of mixing for different cooling strengths (and hence different Damk\"ohler numbers). Second, we check for the effects of including magnetic fields in turbulent box simulations similar to \citet{Gronke2022SurvivalMedium}.

The structure of the paper is as follows. We explain the numerical setups for both TRMLs and turbulent boxes in section \ref{sec:setup}. We present the results from the TRML simulations in section \ref{sec:trml}, and the turbulent boxes in section \ref{sec:turb_results}. Then, we discuss our results in section \ref{sec:disc} and conclude in section \ref{sec:conc}. The visualisations related to this study can be found \href{http://hiteshkishoredas.github.io/research/mhd_multiphase.html}{here}.\footnote{\url{http://hiteshkishoredas.github.io/research/mhd_multiphase.html}}

\section{Numerical setup}
\label{sec:setup}

For our simulations, we use the ATHENA++ code \citep{Stone2020TheSolvers}. We use the default HLLC solver for our hydrodynamic (HD) simulations and the default HLLD solver for our magnetohydrodynamic (MHD) simulations, with Piecewise Linear Method (PLM) applied to primitive variables, second-order Runge-Kutta time integrator, adiabatic EOS and a cartesian geometry. Similar to \citet{Gronke2022SurvivalMedium}, we implemented the Townsend radiative cooling algorithm \citep{Townsend2009AnSimulations} for computing the radiative losses, using a cooling curve at solar metallicity fitted using 40 segments of power laws. We also enforce a temperature floor $T_\mathrm{floor} = 4\times 10^4$ K.

\subsection{Turbulent Radiative Mixing Layer (TRML)}
\label{sec:trml}
Turbulent, radiative mixing layers in the astrophysical context have been investigated in the past \citep[e.g.,][]{Begelman1990TurbulentMedium., Slavin1993TurbulentGalaxies, Kwak2010NUMERICALCALCULATIONS, Hillier2023TheAtmosphere, Fielding2020MultiphaseLayers}.
We use the same numerical setup as the one used in \citet{Tan2021RadiativeCombustion} and \citet{Ji2019SimulationsLayers} to investigate the Turbulent Radiative Mixing Layer (TRML), with a small difference in our coordinates convention. The shear velocity in our simulations is along $\hat{x}$ and the cold/hot interface is normal to $\hat{z}$.

\noindent The different density and velocity profiles are,
\begin{align}
    \rho(z) &= \rho_{\rm cold} + \frac{\rho_{\rm hot} - \rho_{\rm cold}}{2} \bigg[ 1 - \tanh{\frac{z}{a}}  \bigg] \\
    v_x(z) &= \frac{v_{\rm shear}}{2}\bigg[ 1 + \tanh{\frac{z}{a}}  \bigg]\\
    v_z &= \delta v \exp{\frac{-z^2}{a^2}} \sin{k_x x}\sin{k_y y} \\
    B &= \frac{v_{\rm shear} \rho_{\rm hot}^{1/2} }{\Ma} \hat{B}_0
\end{align}
where $k_x$ and $k_y$ are set to $2\pi/L_{\rm box, x}$ and $2\pi/L_{\rm box, y}$ respectively.

\begin{figure*}
	\includegraphics[width=\textwidth]{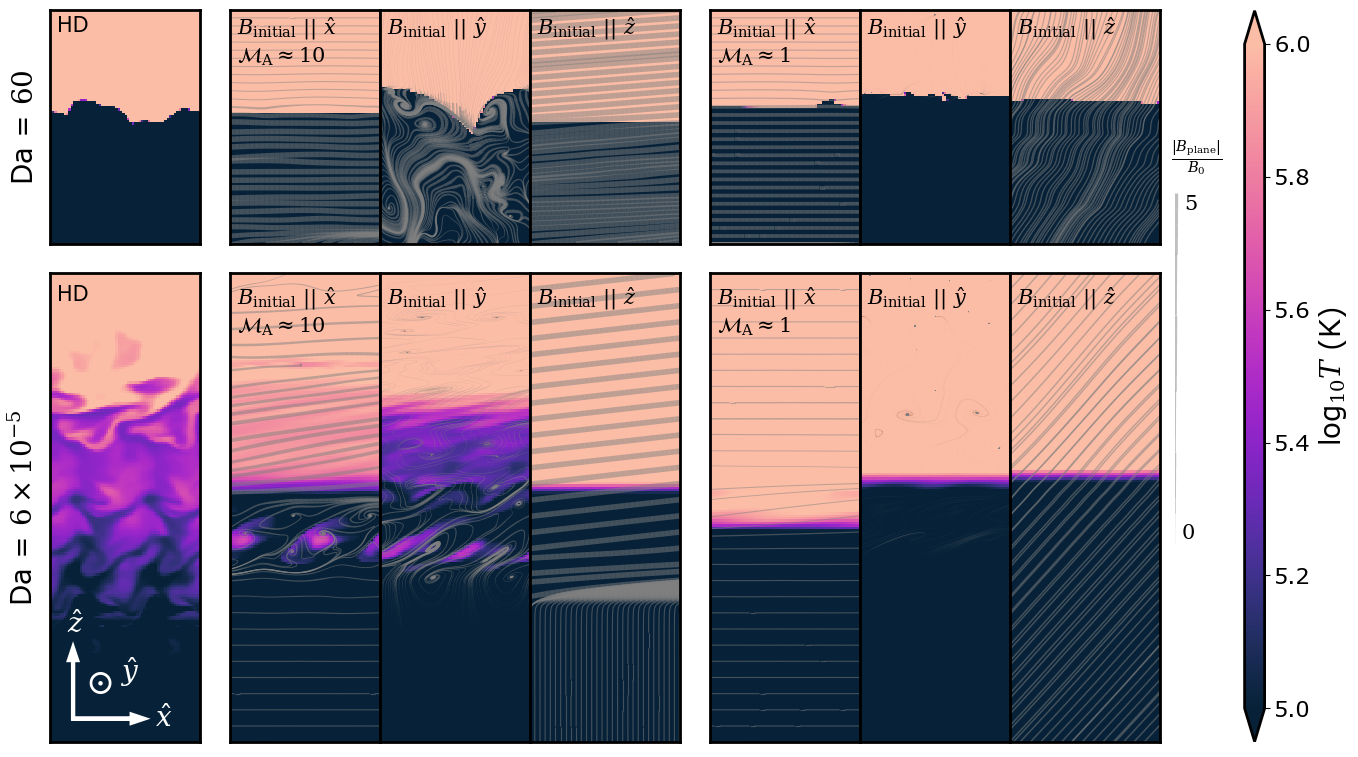}
    \caption{Temperature slices for different TRML simulations for $v_{\rm shear} = 100\kms$ ($\mathcal{M}\approx 0.3$). \textbf{First column} shows the hydrodynamic simulations, \textbf{2$^{\text{nd}}$to 4$^{\text{th}}$ column} show simulations with $\Ma = 10$, \textbf{last three columns} show simulations with $\Ma = 1$. \textbf{Top row} shows simulations with strong cooling, $\Da = 60$, \textbf{Bottom row} shows simulations with weak cooling, $\Da = 6 \times 10^{-5}$. This shows the different ways magnetic fields evolve for different initial orientations. It also suggests that the cases with the higher magnetic field strength along the shear direction show a lesser extent of mixing.}
    \label{fig:T_B_profiles}
\end{figure*}

We use $v_{\rm shear} = 100 \text{km/s}, 50 \text{km/s}$ as the shear velocity (corresponding to sonic Mach numbers of $\mathcal{M}\equiv v_{\rm shear}/c_{\rm s,hot}\sim 0.33, 0.16$), $\delta v = 0.01 v_{\rm shear}$, $a = L_{\rm box,z}/20$ for the interface thickness, and $\Ma = 1, 10$ as initial Alfvénic Mach number. Furthermore, $\rho_{\rm hot} = 1.6 \times 10^{-4} \text{cm}^{-3}$ is the density of hot medium, $\rho_{\rm cold} = 1.6 \times 10^{-2} \text{cm}^{-3}$ is the density of cold medium, and $\hat{B}_0$ is the initial magnetic field direction. We use a floor temperature of $4 \times 10^4$ K, and we stop cooling at temperatures above $0.5 T_{\rm hot}$ to emulate the effect of heating, where $T_{\rm hot}$ is the hot medium temperature. We initialise the cold medium at the floor temperature ($4 \times 10^4~$\textchange{K), }and use a fixed pressure over the whole box to ensure pressure balance in the initial conditions. This corresponds to a hot medium at $T_{\rm hot} = T_{\rm cold} \chi$ with $\chi\equiv \rho_{\rm cold}/\rho_{\rm hot}=100$ being the overdensity. We impose an outflow boundary condition along the normal to the cold/hot interface ($\hat{z}$), and periodic boundary conditions in all other directions. Note that due to the self-similarity of the solution, the chosen numerical values for $\rho$ and $v_{\rm shear}$ are unimportant (as long as the critical dimensionless quantities are kept constant).

We use a resolution of $64 \times 64 \times 640$ in $\hat{x}$, $\hat{y}$ and $\hat{z}$ directions respectively, and use different box sizes to vary parameters in our simulations, but keep the ratios of box lengths in different dimension fixed, $L_{\rm box, z} = 10 L_{\rm box, x} = 10 L_{\rm box, y}$. We vary the Damkohler number \textchange{($\Da = t_{\rm turb}/t_{\rm cool} = L_{\rm box, x~or~ y}/(u_{\rm turb} t_{\rm cool})$)} in a range of $\sim 10^{-4} - 10^4$ by changing $L_{\rm box}$.

In such TRML simulations, the mixing layer tends to move into the hot medium as it is consumed and more cold gas is created. This can cause the mixing layer to go out of the computational domain, especially for high \Da cases. To counter that, we add a velocity to the whole box in the opposite direction to keep the mixing layer inside the computational domain. This velocity is calculated using the difference between the current cold/hot interface position and the original cold/hot interface position ($z=0$). We verify that this does not affect the mixing rates, and only increases the time the mixing layer spends inside the computational box. We denote the different TRML simulations as \texttt{Ma(A)\_Bx(B)}, where \texttt{A} is the Alfvénic Mach number and \texttt{B} is the initial magnetic field orientation.

\subsection{Driven turbulence boxes}

We use a separate simulation setup, similar to the one used in \citet{Gronke2022SurvivalMedium}, to study the behaviour of cold gas in fully-developed turbulence with (MHD) and without (HD) magnetic fields. We start with a box filled with isobaric gas at uniform density and temperature ($T_{\rm hot} = 4 \times 10^6~$K), with solar metallicity and H-abundance. In our MHD simulations, we initialize the box with a uniform magnetic field. We use the Ornstein-Uhlenbeck (OU) process \citep{Eswaran1988AnTurbulence, Schmidt2006NumericalTurbulence} to drive the turbulence at the largest scale ($k=2\pi/L_{\rm box}$), i.e. the box size. We use driving timescale of $0.001~t_\mathrm{eddy}$, correlation timescale $\sim t_\mathrm{eddy}$ and solenoidal to compressive fraction $ f_{\rm sol} =0.3$. We also maintain a $L_{\rm box}/R_{\rm cloud} = 40$ for all our simulations.

We drive the turbulence for 7 $t_{\rm eddy}$ with the cooling turned off, which gives the setup enough time to reach a steady-state with equilibrium kinetic energy and magnetic energy (when included). We restart the simulation after introducing a dense cloud, with an overdensity $\chi = 100$ and radius $R_{\rm cl}$ in the centre of the box while conserving the kinetic, thermal and magnetic energy density. This results in an isobaric, cold, dense cloud with density $\rho_{\rm cold} = \chi \rho_{\rm hot}$ and temperature $T_{\rm cold} = 4 \times 10^4~\mathrm{K} = T_{\rm floor}$. As we use an adiabatic equation-of-state, the average temperature can increase significantly by the end of the turbulent driving phase, due to the dissipation of the turbulent energy. Hence, to bring the ambient temperature back to the required value, before introducing the cloud, we also rescale the internal energy of the whole box by a fixed constant. We also verify that this abrupt rescaling does not have any significant effect on the velocity distribution. 

The input parameters for the driven turbulence are the kinetic energy injection rate ($\dot{E}$), the size of the simulation box ($L_{\rm box}$) and the density of the medium ($\rho$). Given a box size and gas density, we calculate the required $\dot{E}$ for a required turbulent velocity (see, e.g., \citealp{Lemaster2009DISSIPATIONTURBULENCE}), i.e., 

\begin{equation}
      \dot{E} t_{\rm eddy} = (\frac{1}{2} \rho v_{\rm turb}^2 + \frac{1}{2} B_{\rm turb}^2) L_{\rm box}^3 
\end{equation}
and assuming equipartition this yields
\begin{align}
     \dot{E} = \frac{\alpha}{2} \rho L^2 v_{\rm turb}^3
\end{align}
where, $\alpha = 1$ for hydrodynamic simulations and $\alpha = 2$ for MHD simulations \textchange{ } . Following the convention from previous studies, we also use cloud radii normalised by $l_{\rm shatter} = \min{(c_{\rm s} t_{\rm cool})}$. This corresponds to the $c_{\rm s} t_{\rm cool}$ of the cold, dense medium in our simulations, i.e. $c_{\rm s}t_{\rm cool} (\rho_{\rm cold}, T_{\rm floor})$. 

\begin{figure*}
	\includegraphics[width=\textwidth]{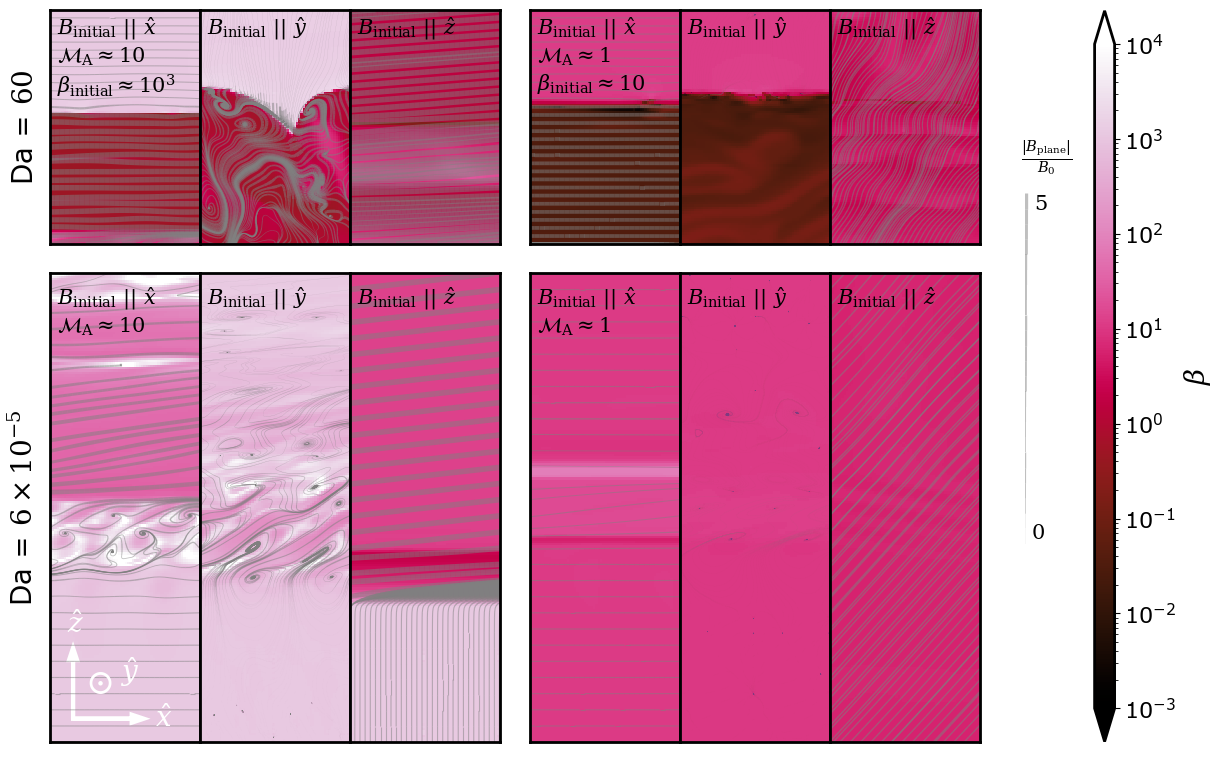}
    \caption{$\beta = P_{\rm thermal}/P_{\rm magnetic}$ slices for different TRML simulations with $v_{\rm shear} = 100 \text{km/s}$ (corresponding to the temperature slices shown in Fig.~\ref{fig:T_B_profiles}). \textbf{1$^{\text{st}}$to 3$^{\text{rd}}$ column} show simulations with $\Ma = 10$, \textbf{last three columns} show simulations with $\Ma = 1$. \textbf{Top row} shows simulations with strong cooling, $Da = 60$, \textbf{Bottom row} shows simulations with weak cooling, $\Da = 6 \times 10^{-5}$. This shows the extent of amplification possible in the different cases. Even with a higher initial $\beta$, turbulent motions can amplify the magnetic fields to lower $\beta$. For cases with lower initial $\beta$ strong cooling in the mixing layer can also lead to amplification.}
    \label{fig:beta_B_profiles}
\end{figure*}


\section{Results: Turbulent Radiative Mixing Layer}
\label{sec:KH_results}

Turbulent Radiative Mixing Layers (TRMLs) are mixing layer simulations that also include radiative cooling for the mixed gas. These have long been studied as an idealised small-scale setup for the mixing between different phases in a multiphase gas \citep[e.g.][]{Esquivel2006MagnetohydrodynamicModels, Ji2019SimulationsLayers, Fielding2020MultiphaseLayers, Tan2021RadiativeCombustion, Yang2023RadiativeNumbers}. In this section, we study the evolution of TRMLs for different cooling strengths and look for differences caused by the presence of magnetic fields.

\begin{figure}
	\includegraphics[width=\columnwidth]{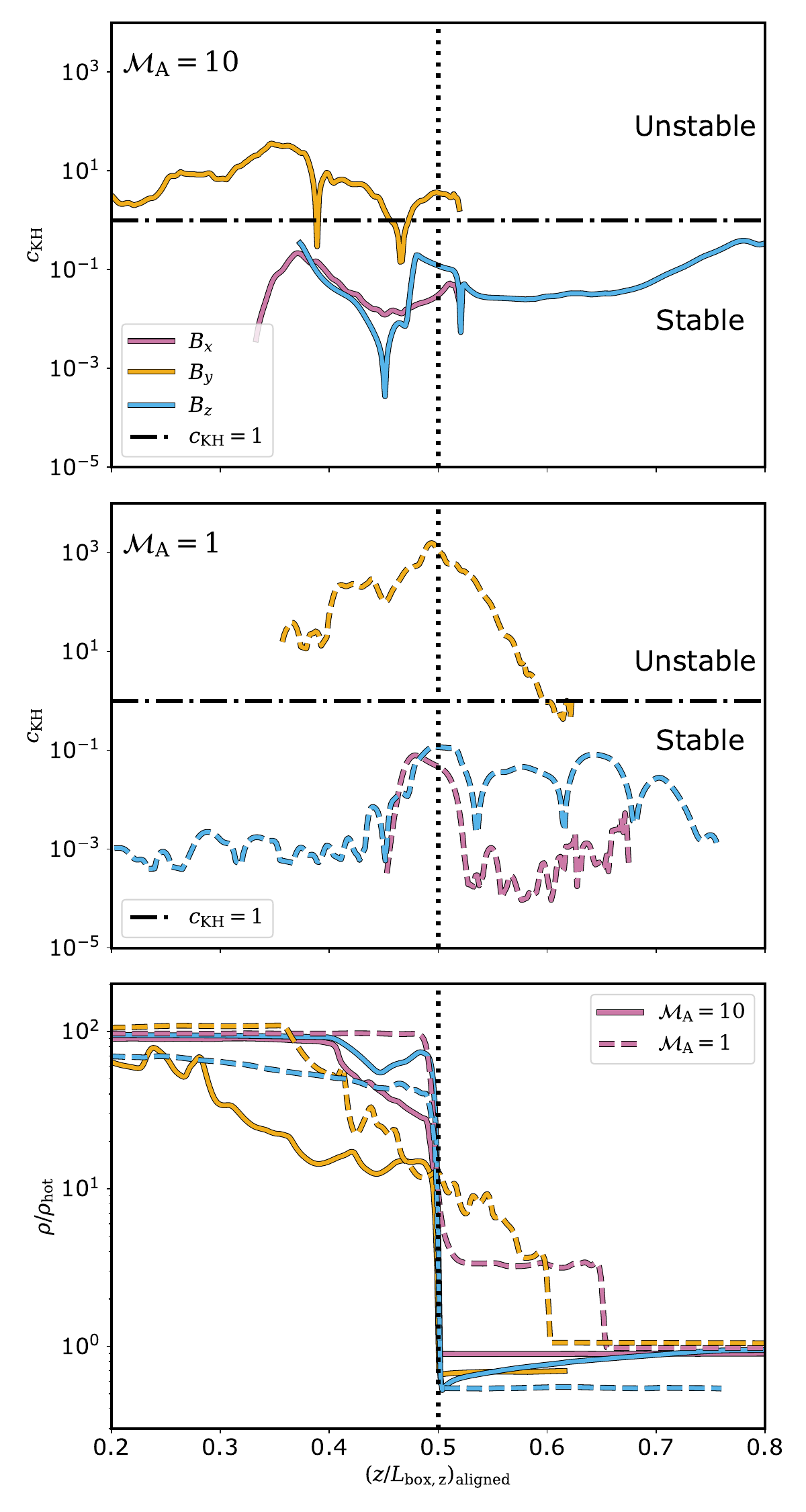}
    \caption{\textbf{Top \& middle} Stability criterion of Kelvin-Helmholtz instability for different initial magnetic field orientations (cf. Eq.~\eqref{eq:KH_cond}), \textbf{Bottom} Density profile for the different cases shown above. This shows the difference in the stability of the mixing layers for different cases. The cases with higher magnetic field strength along the shear (initial or amplified) are more stable. The density profile shows the extent of the mixing layers.}
    \label{fig:KH_cond}
\end{figure}

\begin{figure*}
	\includegraphics[width=\textwidth]{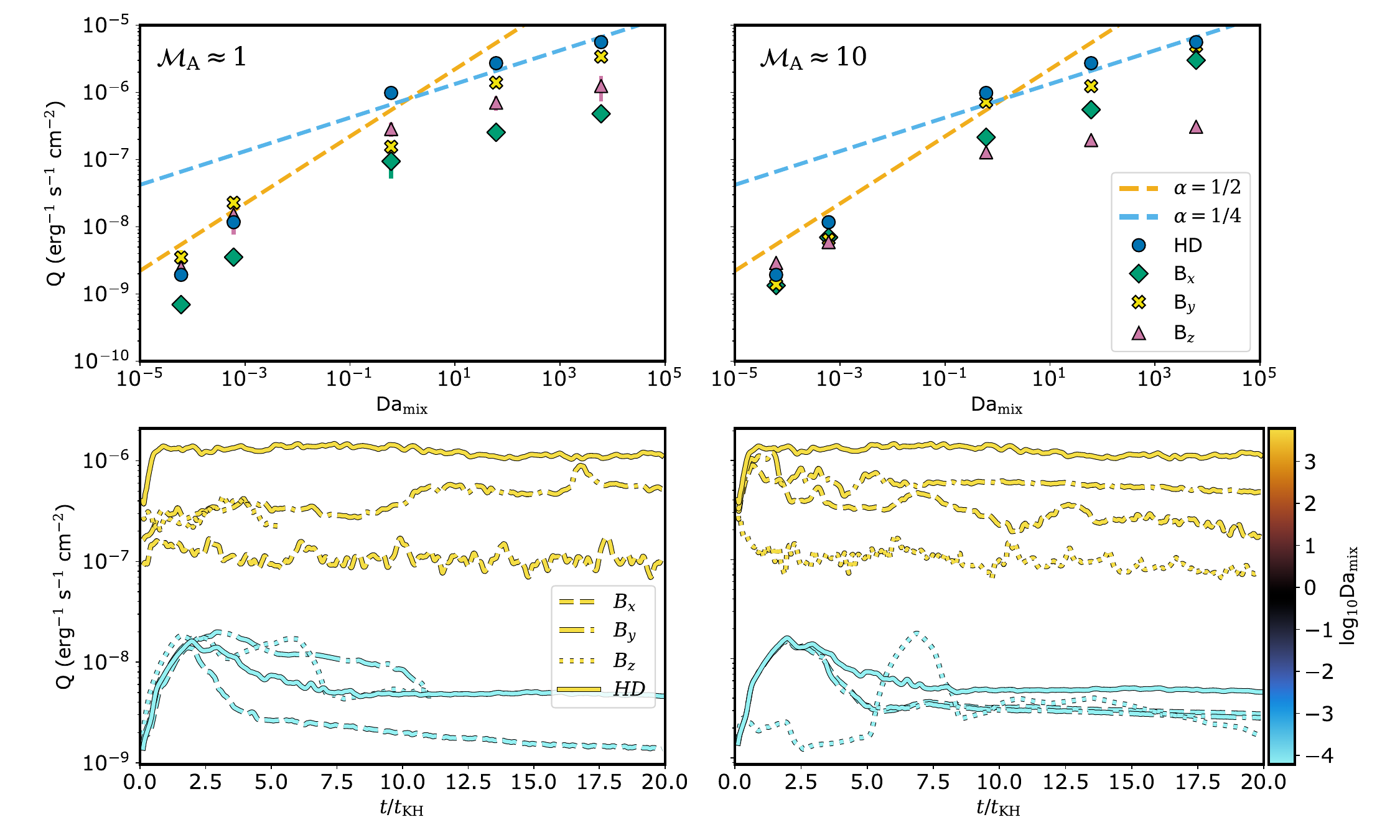}
    \caption{\textbf{Left column} $\Ma = 1$, \textbf{Right column} $\Ma = 10$, \textbf{Top row} Stable values of mixing layer surface brightness for different \Da, The orange and blue dashed lines on the top row panels are the expected values from hydrodynamic TRML simulations by \citet{Tan2021RadiativeCombustion}. This shows clear suppression in cooling rates of most of the simulations with magnetic fields, in comparison to hydrodynamic TRML simulations. We discuss details about the trends in \S~\ref{sec:KH_results}. \textbf{Bottom row} Evolution of mixing layer surface brightness with time, for different initial magnetic field orientations at two \Da values.}
    \label{fig:Q_vs_Da}
\end{figure*}
\subsection{Gas \& magnetic field morphology}

Previous studies of linear and non-linear evolution of mixing layers have shown \textchange{suppression of Kelvin-Helmholtz (KH) instability-induced mixing}, in the presence of magnetic fields. In the linear regime, the KH instability is suppressed for cases with magnetic field along the shear direction if $v_{\rm shear} < (v_{\rm A, hot}v_{\rm A, cold})^{1/2} \approx v_{\rm A, hot} \chi^{-1/4}$  (\cite{Chandrasekhar1961HydrodynamicStability}; Chapter XI, Eq. 205). While, in non-linear regime, the KH instability is suppressed for all magnetic field orientations \citep{Ji2019SimulationsLayers}. In this subsection, we reproduce these results and extend them by varying the cooling and magnetic field strength.

Fig~\ref{fig:T_B_profiles} shows the temperature slices of the different TRML simulations, along with the magnetic field morphology, for different Alfvénic mach number ($\Ma = v_{\rm shear}/v_{A}$, where $v_{A}$ is the Alfvén wave speed), and initial magnetic field orientation ($\hat{B}_0$). We control the $\Da=t_{\rm turb}/t_{\rm cool,mix}$ (where $t_{\rm turb}=L/v_{\rm turb}$ and $t_{\rm cool,mix}$ is the cooling time evaluated at $T=2\times 10^5$ K, and $v_{\rm turb}$ is calculated using scaling relations from \citealp{Tan2021RadiativeCombustion}; we will investigate the role of $v_{\rm turb}$ more below) by varying the box sizes, and \Ma by changing the initial magnetic field strength, for a given sonic Mach number. We have simulations with $\mathcal{M}_{\rm shear, hot}$ fixed to 0.16 and 0.33, corresponding to a $v_{\rm shear}$ of 50 and 100 \kms. We find that the amplification in the magnetic fields is very different depending on the initial magnetic field orientation as we discuss below.

Fig.~\ref{fig:beta_B_profiles} shows the $\beta(=P_{\rm thermal}/P_{\rm magnetic})$ and has layout similar to that of Fig.~\ref{fig:T_B_profiles}. The extent of amplification in the different cases is much clearer in Fig.~\ref{fig:beta_B_profiles}, where the darker regions correspond to a higher magnetic field strength and lighter regions to a lower magnetic field strength.

The upper and lower row of Fig.~\ref{fig:T_B_profiles} shows the fast and slow cooling cases, respectively. On one hand, in the fast cooling ($\Da>1$) a sharp temperature edge between $T<10^5\,$K and $T\sim 10^6\,$K gas is visible, i.e., a true multiphase structure exists, while on the other hand, a large amount of this ``intermediate temperature'' gas is visible in the slow cooling ($\Da < 1$) case -- with the exact amount depending on the suppression of mixing caused by magnetic fields. Unsurprisingly, in the HD case most mixed gas exists and generally in the \textchange{MHD case with} $\Ma\approx 1$ the least. What is maybe a bit more surprising is the effect (and the evolution of) the magnetic field topology: in the $\mathbf{B}\parallel \hat x$ case the suppression of mixing is easy to understand and expected from linear theory \citep{Chandrasekhar1961HydrodynamicStability}. However, we also find in all the other cases a (strong) suppression. $\mathbf{B}\parallel \hat y$ also has a strong effect, particularly in the $\Ma\approx 1$ simulations. This is due to the amplification of the magnetic fields in the direction of the shear, as discussed below. On the other hand, for $\mathbf{B}\parallel \hat z$ initially, one can note two distinct effects depending on the value of \Ma. For $\Ma > 1$ the flow bends the magnetic field lines, resulting in a similar situation as in the $\mathbf{B}\parallel \hat x$ case; in fact, even larger suppression since the bending of the field lines leads to a $B_x > B_{\rm initial}$. An artefact of this bending can be seen in magnetic field topology in the bottom panel for $\Ma>1$ with $\mathbf{B}\parallel \hat z$. We find a kink in the magnetic field moving downwards at the Alfvén wave speed. For $\Ma \lesssim 1$, however, the magnetic field lines are so stiff that a bending by $90$ degrees is not possible. Instead, we end up with diagonal field lines which, nevertheless, substantially suppress the mixing. 

To better understand the exact order of amplification, we first consider the cases where the shear is super-Alfvénic ($\Ma\sim 10$) in the hot medium, in the three central (2$^{\rm nd}$ - 4$^{\rm th}$) columns of Fig.~\ref{fig:T_B_profiles} \& \ref{fig:beta_B_profiles}, to explain the extent of amplification in the different cases.

\begin{itemize}
    \item $B_{\rm initial}~||~\hat{z}~||~\hat{n}_{\rm interface}$ (\texttt{Ma10\_Bz}): The amplification is the highest for this case. The Alfvén wave velocity in the dense medium is lower by a factor of $\chi^{1/2}$, \textchange{hence the field lines are more ``anchored'' in the cold gas, compared to the hot gas. This causes field lines to bunch up near the interface, and result} in high amplification of magnetic fields in the direction parallel to the shear. This amplification is so high that the magnetic field strength can get much higher than the initialised magnetic field strength.
    \item For $B_{\rm initial}~||~\hat{x}~||~\vec{v}_{\rm shear}$ (\texttt{Ma10\_Bx}): \textchange{As the Kelvin-Helmholtz (KH) instability grows, it gives rise to vortices around $\hat{y}$. Fig.~\ref{fig:T_B_profiles} shows that vortices can stretch and bend the magnetic fields, leading to their amplification. These vortices can further become turbulent and cause more amplification due to local dynamo effect. All of these put together, result in the second-highest amplification of magnetic fields along the shear direction.}
    \item Lastly, for $B_{\rm initial}~||~\hat{y}~\perp~\vec{v}_{\rm shear}~\perp~\hat{n}_{\rm interface}$ (\texttt{Ma10\_By}): The amplification is the lowest \textchange{as the only process for amplification of magnetic fields is due dynamo effect from the turbulent motions generated in the mixing layer due to non-linear evolution of the KH instability.}
\end{itemize}
This results in a general order for the magnetic field strength along the shear direction in Super-Alfvénic TRML simulations as: \texttt{Ma10\_Bz} $>$ \texttt{Ma10\_Bx} $>$ \texttt{Ma10\_By}.
 
\noindent Similarly, for the cases where the shear is sub/trans-Alfvénic ($\Ma\sim~1$) in the hot medium, in the three rightmost columns in Fig.~\ref{fig:T_B_profiles}, we again check for magnetic field strength along the shear direction.
\begin{itemize}
    \item For $B_{\rm initial}~||~\hat{z}~||~\hat{n}_{\rm interface}$ (\texttt{Ma1\_Bz}): As the shear is sub-Alfvénic, the amplified magnetic field in the shear direction is not high enough to surpass the initialised magnetic field in the shear direction for $B_{\rm initial}~||~\hat{x}~||~\vec{v}_{\rm shear}$ (\texttt{Ma1\_Bx}). So, \texttt{Ma1\_Bz} ends up with the second highest in the order of magnetic field strength along the shear direction.
    \item $B_{\rm initial}~||~\hat{x}~||~\vec{v}_{\rm shear}$ (\texttt{Ma1\_Bx}) has the strongest magnetic field in shear direction, just due to the high initial magnetic strength.
    \item This leaves the $B_{\rm initial}~||~\hat{y}~\perp~\vec{v}_{\rm shear}~\perp~\hat{n}_{\rm interface}$ (\texttt{Ma1\_By}): Due to the much higher overall magnetic field strength, it is harder for the resulting turbulent velocity to cause any amplification.
\end{itemize}
Hence, we get an order for the magnetic field strength along the shear direction in Sub-Alfvénic TRML simulations as: \texttt{Ma1\_Bx} $>$ \texttt{Ma1\_Bz} $>$ \texttt{Ma1\_By}. In our simulations, we find one exception to this order, at intermediate $\Da$, where the order of \texttt{Ma1\_Bz} and \texttt{Ma1\_By} is switched. 

\begin{figure*}
	\includegraphics[width=\textwidth]{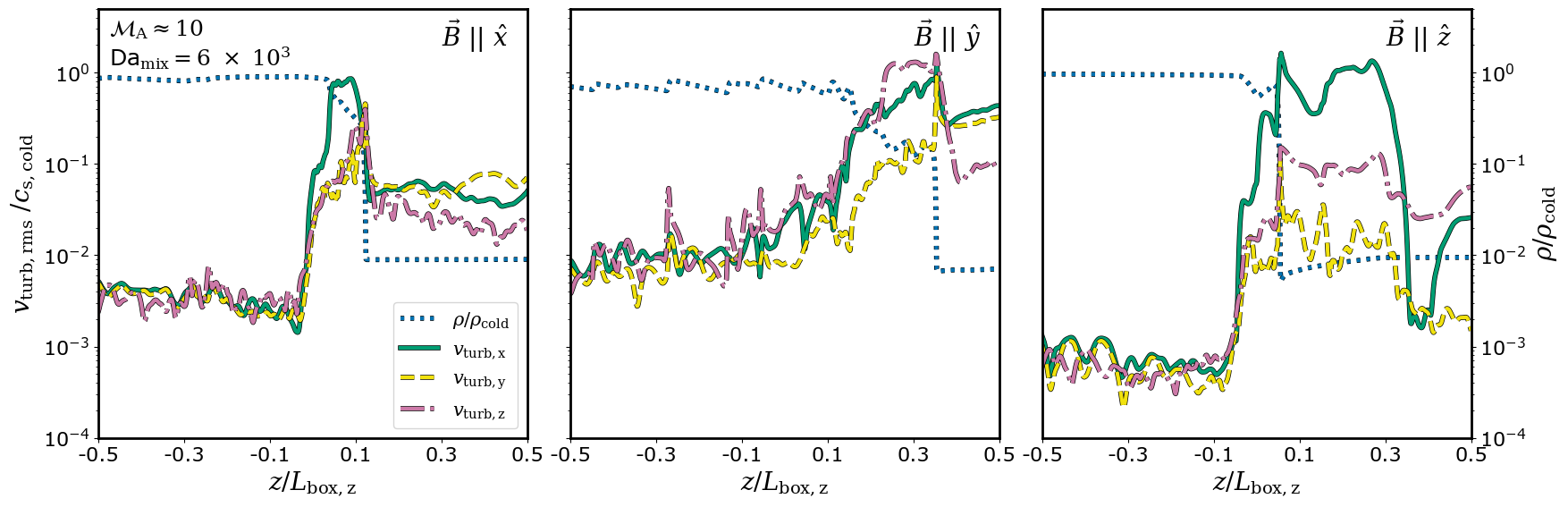}
    \caption{$u^\prime$ profiles along $\hat{z}$ for different initial magnetic field orientations. The first two panels from the left show the difference between the direction parallel to magnetic fields versus the other directions. The third panel shows the outlier case of magnetic fields normal to the interface, where both the normal ($\hat{z}$) and shear direction show much higher fluctuations due to the presence of magnetic fields along both these directions. Hence, we choose the directions which free from these spurious fluctuations in these different cases, as denoted in Eq~\ref{eq:uprime}}
    \label{fig:uprime_profiles}
\end{figure*}

\begin{figure}
	\includegraphics[width=\linewidth]{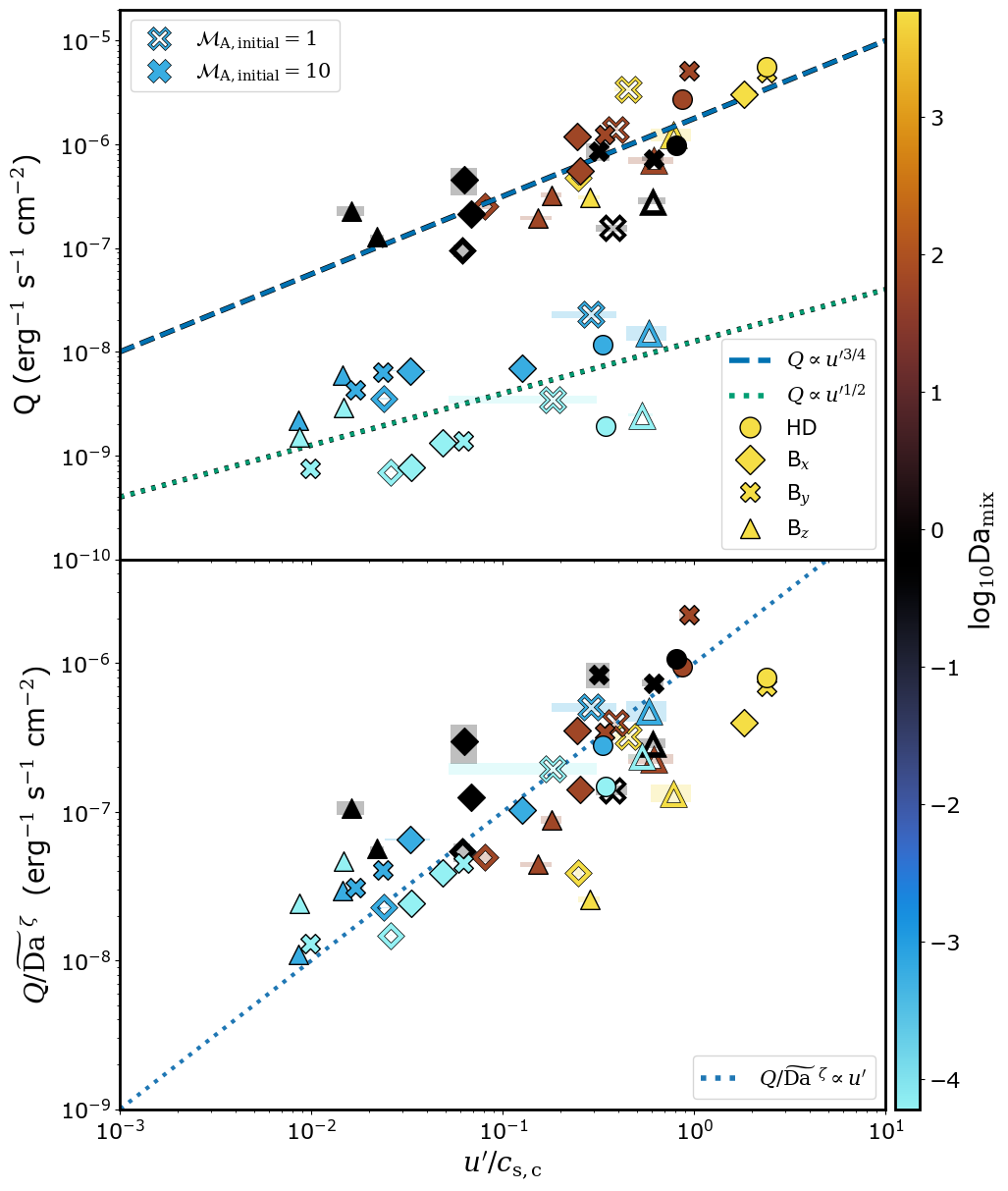}
    \caption{\textbf{Top panel} Scatter plot of the surface brightness ($Q$) and turbulent velocity ($u^\prime$) calculated from the simulations. The dashed and dotted lines show the respective strong and weak cooling scaling relation according to Eq.~\ref{eq:Q_Da_strong}-\ref{eq:Q_Da_weak}. \textbf{Bottom panel} Similar to the top panel, after we remove the $\widetilde{\Da}$ dependence. The dotted line shows the analytical expectation from \citet{Tan2021RadiativeCombustion}, which they find for hydrodynamic simulations. This suggests that the general relation found in hydrodynamic TRML simulations, between the turbulent velocity in the mixing layer and cooling (and hence mixing) rate, is still valid in the presence of magnetic fields.}
    \label{fig:Q_vs_vturb}
\end{figure}

\subsection{Cooling rates}

According to the linear KH instability criterion, the stronger the magnetic field in the direction of the shear, the more stable the perturbation gets. This means stronger magnetic fields in the later non-linear phase may disrupt the generation or cascade of further vortices. To test this, we quantify the stability of KH perturbations, using the linear stability criterion \citep{Chandrasekhar1961HydrodynamicStability}. So, the perturbation is stable to KH instability, if
\begin{align}
    v_{\rm shear} &< (v_{\rm A, hot}v_{\rm A, cold})^{1/2} = v_{\rm A, hot}\chi^{-1/4}
    \end{align}
    which can be expressed as a dimensionless number
    \begin{align}
    c_{\rm KH} = \frac{\Delta v_{x} \chi^{1/4} \rho_{\rm hot}^{1/2}  }{B_x}  <  1.
    \label{eq:KH_cond}
\end{align}
We calculate $c_{\rm KH}$ using profiles of all the relevant quantities along the normal to the hot/cold interface. This results in a profile of $c_{\rm KH}$, in which a value <1 denotes a tendency towards stability while a value >1 shows a tendency towards instability. Fig.~\ref{fig:KH_cond} shows the profiles of this KH stability criterion at an advanced stage of evolution, for different magnetic field orientations and strengths. The top and middle panels show the KH stability criterion and the bottom shows the corresponding density profiles. We align the profiles so that the hot/cold interface aligns between all the cases. We do not plot the points on the profile which have a $\Delta v<10^{-2}$ km/s, or if the points are out of the computational domain.

We find that due to the amplification of the magnetic fields in the shear direction, the KH instability is suppressed. The order of the extent of suppression seems to follow the same order as the amplification of the magnetic fields. For both $\Ma>1$ and $\Ma<1$, \texttt{Ma1\_By} and \texttt{Ma10\_By} are the most unsuppressed as the $c_{\rm KH}$ is almost entirely in the unstable regime. For the other two directions, the extent of suppression depends on the size of the portion around the mixing layer that is stable. This means, \texttt{Ma1\_Bx} and \texttt{Ma10\_Bz} are the most suppressed in $\Ma<1$ and $>1$, respectively. This trend in suppression is important as this can affect the cooling rates, which we check next.

We study the cooling rates using the surface brightness of TRML simulations for all the different cases, that is, different values of the Dahmköhler number ($\Da = t_{\rm turb}/t_{\rm cool}$), Alfvénic mach number ($\Ma = v/v_{A}$, where $v_{A}$ is the Alfvén wave speed), and initial magnetic field orientation ($\hat{B}_0$). 
We define the surface brightness as the total luminosity per unit surface area, that is, $Q = L_{\rm total}/(L_x L_y)$. Fig.~\ref{fig:Q_vs_Da} shows the calculated saturation surface brightness (along with $2\sigma$ errorbars) for the different simulations on the top row, along with its temporal evolution on the bottom row. 

We find that the order of amplification of magnetic fields along the shear direction, mentioned above, matches the order of suppression of surface brightness, as shown in the right panel in Fig.~\ref{fig:Q_vs_Da}. This is due to the higher suppression of KH instability by the higher magnetic field strength along the shear direction, as also expected from the linear theory \citep{Chandrasekhar1961HydrodynamicStability}. Below, we dive deeper into this correlation.

We also find that the difference in cooling rates, due to this suppression of KH instability, is reduced for the low \Da cases with $\Ma\sim10$, as shown in Fig.~\ref{fig:Q_vs_Da}. This might be due to the change in the rate-limiting process. For a low \Da the cooling is very slow, so the cooling rate is bottlenecked by the slow cooling rate rather than the mixing rate. This does not happen for $\Ma\sim1$ because the mixing rate is suppressed to such low values that mixing continues to be the rate-limiting process.

\begin{figure*}
	\includegraphics[width=\textwidth]{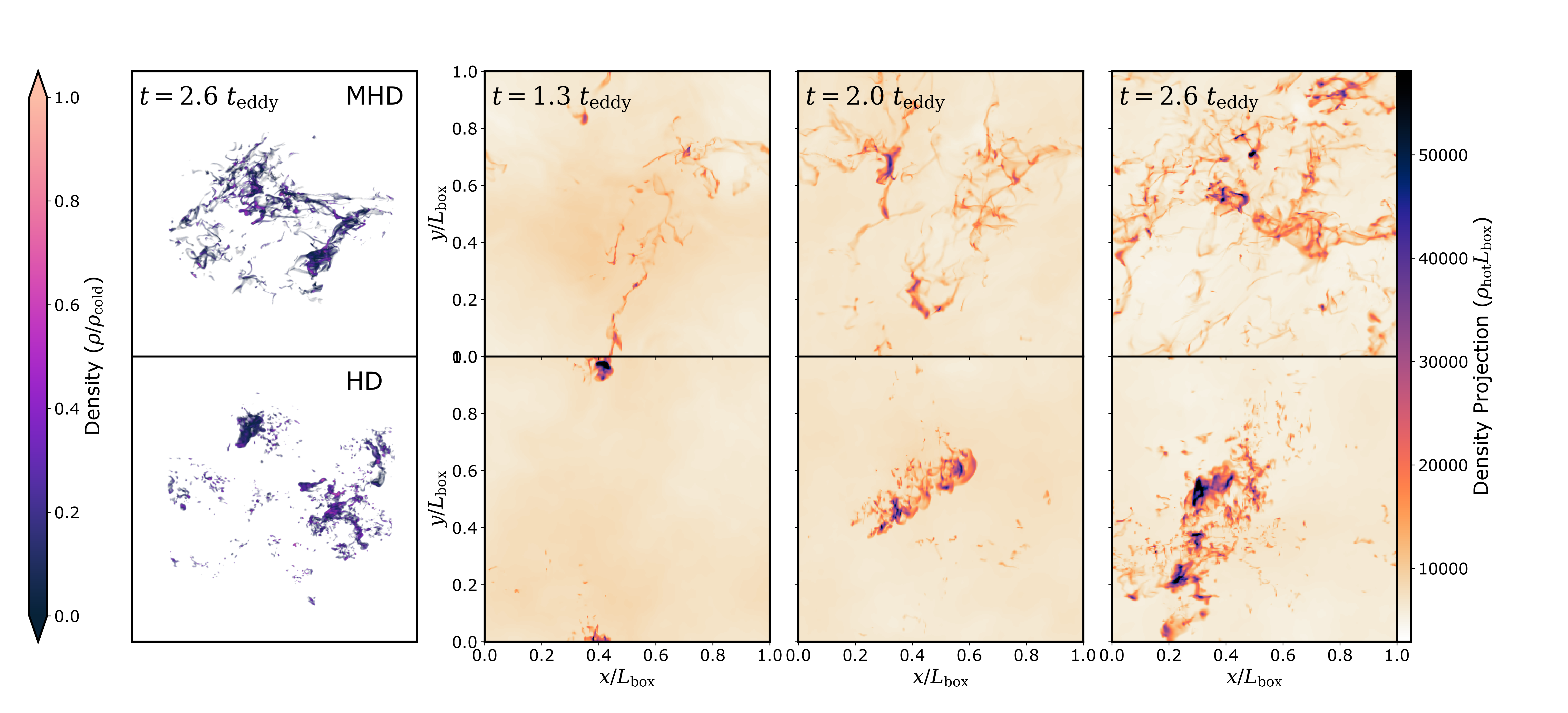}
    \caption{\textbf{Left column} Density rendering at $2.6 t_{\rm eddy}$ after the cold gas cloud of size $310 l_{\rm shatter}$ is introduced in a turbulent box with rms velocity of Mach 0.5, \textbf{2\textsuperscript{nd}-4\textsuperscript{th} column} Density projections of the same simulation, at $1.3$, $2.0$ and $2.6 t_{\rm eddy}$ after the cold gas cloud is added. The \textbf{top row} panels are from the simulations with magnetic fields, and the \textbf{bottom row} panels are from the simulation without magnetic fields. These show the clear differences between the morphology of cold gas of gas with and without magnetic fields, while also showing the similarities in the overall evolution of the cold gas.}
    \label{fig:turb_snapshot}
\end{figure*}

\subsection{Turbulence velocity profiles}

Among the different ways to mix two phases, turbulence is one of the most efficient ones. In this section, we quantify the extent of turbulence in the mixing layer in the above discussed TRML simulations and look for connections of turbulence with the rate of mixing and cooling in the system.

This dependence has been studied before, in the absence of magnetic fields. Assuming a constant pressure and cooling function for hydrodynamic TRML simulations, as shown in \citet{Tan2021RadiativeCombustion}, for a strong cooling regime ($\Da>0.1$),
\begin{align}
    Q \propto u^{\prime 3/4} L_{\rm box}^{1/4} \propto u^{\prime} \Da^{1/4}  \label{eq:Q_Da_strong}
\end{align}
and for slow cooling regime ($\Da<0.1$),
\begin{align}
    Q &\propto u^{\prime 1/2} L_{\rm box}^{1/2} \propto u^{\prime} \Da^{1/2}.  \label{eq:Q_Da_weak}
\end{align}

Our next step is to check these relations in the presence of magnetic fields.
We use the geometric method to calculate the $u^\prime$, similar to \citet{Tan2021RadiativeCombustion}. First, we calculate this bulk velocity profile as the density-weighted average of the velocity field along the other two perpendicular axes (i.e. $\hat{x}$ and $\hat{y}$). Then, we subtract the bulk velocity profile along $\hat{z}$, which is normal to the hot/cold interface, from the velocity field to obtain the turbulent component. This turbulent velocity field allows us to compute density-weighted RMS values of this field along the axes perpendicular to the hot/cold interface normal (i.e. $\hat{x}$ and $\hat{y}$), to obtain 1D profile of all three components of turbulent velocity along normal to the hot/cold interface ($\hat{z}$). Fig.~\ref{fig:uprime_profiles} shows an example of the calculated turbulent velocity profiles for a snapshot where the cooling rate has reached saturation, for different initial magnetic field orientations. There are other methods like Gaussian filtering \citep[e.g.][]{1994PhFl....6.1775B,2000ExFl...29..275A,Abruzzo2022TamingInteractions} to get these 1D profiles, but we find that the choice of method does not significantly influence the next steps, as shown in appendix \ref{sec:uprime_method_appendix}.

Unlike \citet{Tan2021RadiativeCombustion}, we cannot select a particular direction that is untouched by shear or cooling inflow, as that direction can be aligned with the magnetic field. Hence, we have to calculate the turbulent velocity using different directions for different cases. In addition to that, the turbulent velocity component along the magnetic field can have some contributions from large velocity fluctuations along the magnetic fields, as Fig~\ref{fig:uprime_profiles} shows. To get around this issue, we calculate the $u^\prime$ using the other two components perpendicular to magnetic fields, except in the case of $B_{\rm initial}~||~\hat{z}~||~\hat{n}_{\rm interface}$ (\texttt{Ma1\_Bz}, \texttt{Ma10\_Bz}). For the exceptions, where the large-scale magnetic field orientations are along the shear and normal to the interface (i.e. $\hat{x}$ and $\hat{z}$), we only consider the turbulent velocity component along the direction perpendicular to those directions, $\hat{y}$. In short, we use the following expressions to calculate the turbulent velocities in the different cases,
\begin{align}
    u^{\prime 2}_{\rm \hat{B}_0||\hat{x}} (z) &= \frac{3}{2}~\bigg( \langle u^\prime_y\rangle^2_{\rm rms} + \langle u^\prime_z\rangle^2_{\rm rms} \bigg)  \\
    u^{\prime 2}_{\rm \hat{B}_0||\hat{y}} (z) &= \frac{3}{2}~\bigg( \langle u^\prime_x\rangle^2_{\rm rms} + \langle u^\prime_z\rangle^2_{\rm rms} \bigg)  \\
    u^{\prime 2}_{\rm \hat{B}_0||\hat{z}} (z) &= 3~ \langle u^\prime_y\rangle^2_{\rm rms} \label{}  \label{eq:uprime}
\end{align}
We, furthermore, checked that the turbulent velocity components used in the equations above are the ones that have similar profiles among themselves in each case, to ensure the isotropicity of the turbulent components, as shown in Fig.~\ref{fig:uprime_profiles}.

As in \citet{Tan2021RadiativeCombustion}, we consider the maximum of the obtained turbulent velocity profile as the turbulent velocity. We repeat this process for every snapshot and consider the mean of the turbulent velocity over the last $5t_{KH}$ as the saturated turbulent velocity ($u^\prime$) and the standard deviation as the error.

Note that throughout we denote the Damk\"ohler number with $\widetilde{\Da}$ when the measured $u'$ was used and $\Da$ when the theoretically expected $u'$ from the \citet{Tan2021RadiativeCombustion} was used.

\subsection{Turbulence vs. cooling rates}


We use the above obtained $u^\prime$ to check the relations in Eq.~\eqref{eq:Q_Da_strong}-\eqref{eq:Q_Da_weak}, as shown in Fig.~\ref{fig:Q_vs_vturb}. The top panel of Fig.~\ref{fig:Q_vs_vturb} shows a scatter plot of the surface brightness and turbulent velocities calculated from the simulations. We find the respective strong and weak cooling scaling relation according to Eq.~\eqref{eq:Q_Da_strong}-\eqref{eq:Q_Da_weak}, regardless of the magnetic field strength and orientation. For a better comparison, we remove the $\widetilde{\Da}$ dependence and show the correlation in the bottom panel of Fig.~\ref{fig:Q_vs_vturb}. We clearly show that regardless of the initial magnetic field orientation or strength, the Eq.~\ref{eq:Q_Da_strong}-\ref{eq:Q_Da_weak} holds true, even though the relations originally obtained for hydrodynamic systems \citep{Tan2021RadiativeCombustion}. 
We also confirm that the method of $u^\prime$ calculation does not affect these results, as shown in Appendix~\ref{sec:uprime_method_appendix}, in Fig.~\ref{fig:Q_vs_vturb_gauss}.



\section{Results: Turbulent box} 
\label{sec:turb_results}
In the previous section, we examined the TRML setup which is considered a more idealized version of Turbulent boxes (cf. \S\ref{sec:setup}), and found that magnetic fields can suppress the mixing in general, regardless of their initial orientation. If we follow this conclusion, one would expect the inclusion of magnetic fields in turbulent boxes to cause significant differences in the evolution of a cold cloud. This effect can manifest either as a change in the cold gas growth rates or a change in the survival criteria.
In this section, we show results from the ``turbulent box'' setup in which we place a cold gas clump of size $R_{\rm cl}$ in a turbulent medium, with a turbulent Mach number of $\mathcal{M}_\mathrm{s}$, either with (MHD) or without (HD) magnetic fields. We then look for the effect of the magnetic fields, not only on the growth rates and survival criteria but also on the morphology and overall behaviour of the cold gas.

\subsection{Cold gas survival and growth}
\label{sec:cold_gas}

When cold gas is subject to turbulence it can either be mixed away in the hot material or the mixed gas cools sufficiently fast to ensure continuous survival of the cold gas. \citet{Gronke2022SurvivalMedium} studied this effect using hydrodynamical simulations and found a relation between the critical value of $\teddy/\tcoolmix$ (equivalently $R_{\rm cl}/l_{\rm shatter}$) for a given turbulent velocity. However, as we have shown in the last section, magnetic fields can suppress \textchange{Kelvin-Helmholtz (KH) instability-induced mixing} via , between the hot and cold phases in a TRML. Hence, one could expect a similar significant difference in a turbulent box with (MHD) and without (MHD) magnetic fields.

As mentioned in (cf. \S\ref{sec:setup}), the initial seed magnetic fields are such that plasma $\beta~(=P_{\rm thermal}/P_{\rm magnetic}) \approx 100$. But, due to the local dynamo effect \citep{Schekochihin2001StructureTheory}, the magnetic field gets amplified to reach equipartition with the turbulent kinetic energy by the end of the driving phase of the turbulence and before the cloud is introduced. We can use the fact that at equipartition, $\Ma\sim 1$, and the relation between the sonic ($\mathcal{M}_\mathrm{s}$) and Alfvénic (\Ma) Mach numbers as follows, to get an estimate of the final plasma $\beta$ for a given $\mathcal{M}_\mathrm{s}$,
\begin{align}
    \beta \sim \frac{2}{\gamma \mathcal{M}_\mathrm{s}^2} \sim \frac{1}{\mathcal{M}_\mathrm{s}^2} .
\end{align}
This means the plasma $\beta$ at equipartition, when the cloud is introduced, is $\approx$ 16, 4 and 1 for $\mathcal{M}_\mathrm{s}=$ 0.25, 0.5 and 0.9, respectively.

\begin{figure}
	\includegraphics[width=\linewidth]{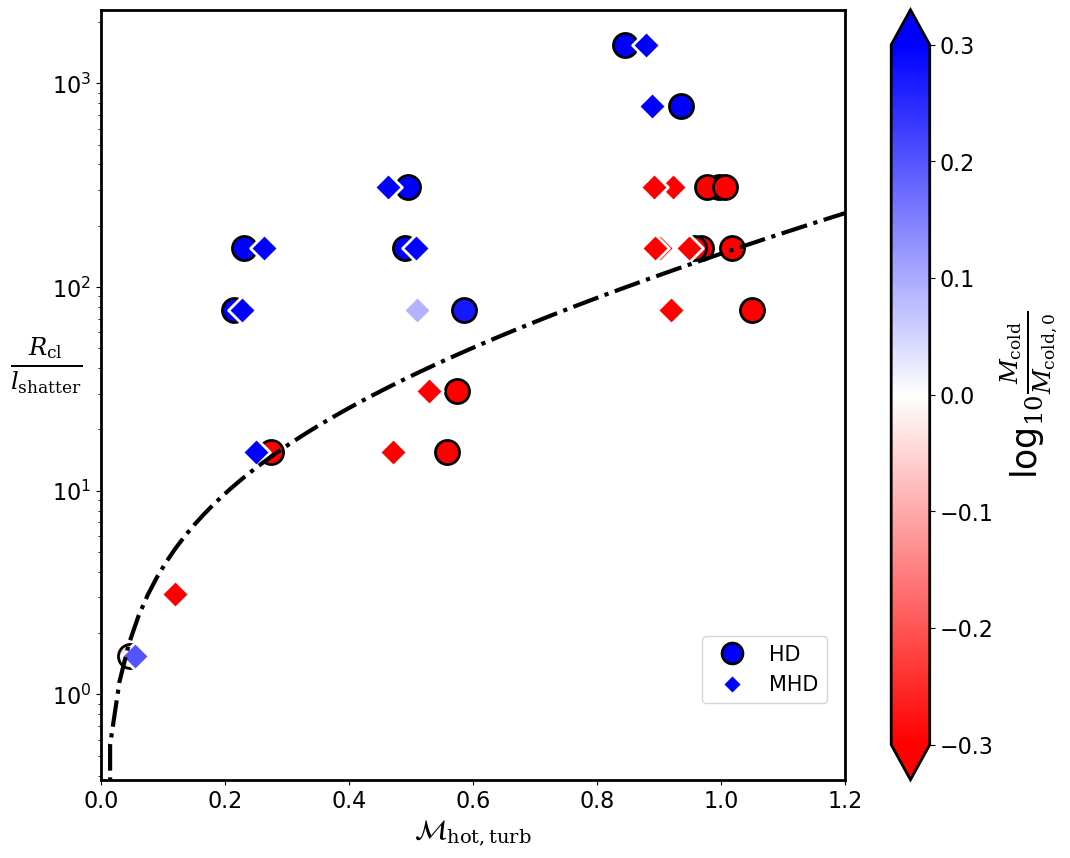}
    \caption{Survival or destruction of the cold gas in the different turbulent boxes. The dashed line is the survival criterion from \citet{Gronke2022SurvivalMedium}. This shows the surprising lack of difference between the survival criterion, with and without magnetic fields. The subsonic turbulent simulations agree well with the previously found survival criterion, with some deviation in trans-sonic turbulent boxes (c.f. \S~\ref{sec:turb_results}).}
    \label{fig:cloud_survival}
\end{figure}

\begin{figure}
	\includegraphics[width=\linewidth]{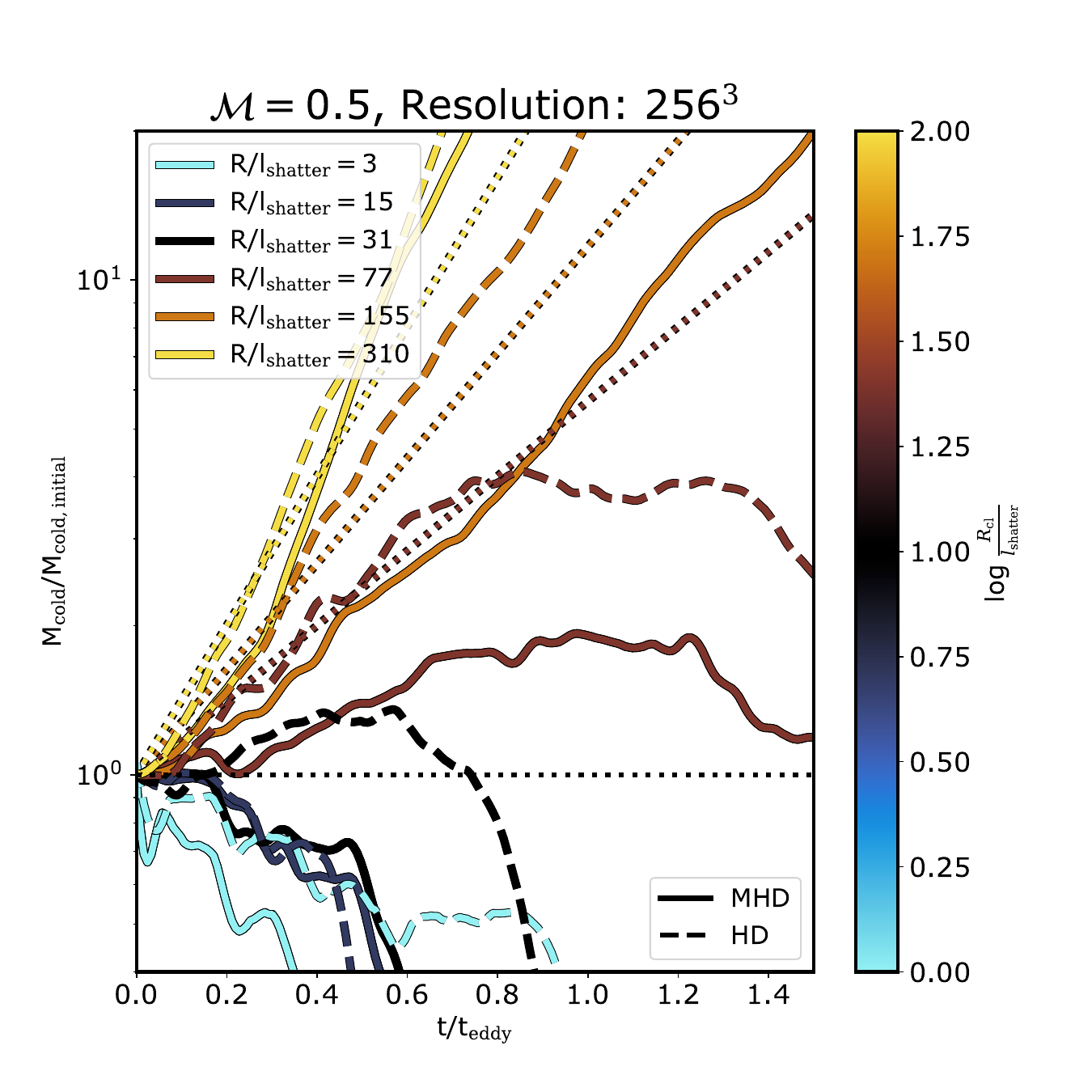}
    \caption{Cold gas evolution with time for different simulations initiated with varying sizes of cold gas cloud in turbulence with $\mathcal{M} = 0.5$. Solid lines show the simulations with magnetic fields, dashed lines show the hydrodynamic simulations and the dotted lines show the expected hydrodynamic growth rates from \citet{Gronke2022SurvivalMedium}. This shows that there are only marginal differences between the growth and destruction rates of the cases with and without magnetic fields, compared to the differences seen in the TRML simulations. The differences further diminish as we consider cases well within the survival regime.}
    \label{fig:cold_gas}
\end{figure}

We study multiple of these turbulent box simulations at different turbulent velocities ($\mathcal{M}_\mathrm{s} \approx ~$ 0.25, 0.5 and 0.9) and with multiple cloud radii near the critical radii found in previous hydrodynamic studies \citep{Gronke2022SurvivalMedium}. See \S\ref{sec:setup} for an overview of the setup. 

Fig.~\ref{fig:turb_snapshot} shows one of the HD-MHD pairs of simulations with $\mathcal{M}_\mathrm{s} \approx 0.5$ and $R_{\rm cloud} \approx 310 l_{\rm shatter}$, where the upper row shows the simulation with magnetic fields (MHD) and the lower row shows the same simulation but without magnetic fields (HD). The leftmost column shows renderings of the number density with the view in the direction of one of the diagonals of the computational domain, at a time $2.6~\teddy$ after the cloud is introduced. The three columns on the right show density projections of the same simulations as the first column, but at different times ($1.3~\teddy$, $2.0~\teddy$ and $2.6~\teddy$ after the cloud insertion). We can see that in both the HD and the MHD simulation the crude behaviour of the cold gas is similar. The cloud survives and grows in both simulations, and the overall amount of gas in the simulation box also looks roughly similar. 

Fig.~\ref{fig:turb_snapshot} also shows how the gas structure evolves. The cold gas seems to grow as it gets stretched, folded and transported by the turbulent motion in the hotter surrounding medium. We also see the difference in the general morphology of the cold gas in the two cases. The cold gas morphology is much more filamentary in the MHD simulation, while it is very clumpy and less dispersed in the HD simulation. We will discuss the morphology further in \S~\ref{sec:cold_morph}.

Next, we check for the growth (or destruction) of the cold gas in all the turbulent box simulations. We define the cold gas mass as the total gas mass with temperature below $2~T_{\rm floor} = 8\times 10^4$ K and normalize the obtained value with the initial cold gas cloud mass. We take the obtained cold gas mass and check for survival or destruction at different Mach numbers and cloud radii. We determine the survival or destruction of the cloud using the final normalized cold gas mass values. The cases with final $M_{\rm cold}/M_{\rm cold,0}>1 (<1)$ are considered to show cold gas survival (destruction). We plot this survival or destruction for the different cases as a scatter plot in Fig.~\ref{fig:cloud_survival}. It clearly shows the lack of difference in survival criteria between the pairs of simulations with (MHD) and without (HD) magnetic fields. This shows that the inclusion of magnetic fields does not affect the survival or destruction of the cold cloud. We also plot the survival criteria found by \citet{Gronke2022SurvivalMedium} in Fig.~\ref{fig:cloud_survival}, given as 

\begin{align*}
    \frac{R_\mathrm{cl}}{l_\mathrm{shatter}} = \mathcal{M}_\mathrm{hot, turb} \frac{t_\mathrm{ cool, mix}}{t_\mathrm{cool, cold}} 10^{(0.6 \mathcal{M}_\mathrm{ hot, turb} + 0.02) }
\end{align*}

\noindent We find that this survival criterion works well for our subsonic turbulent simulations, but our transonic turbulent simulations seem to deviate slightly from this survival criteria. This could be due to the difference between subsonic, transonic and supersonic turbulence due to the presence of shocks in later cases, possibly destroying the clouds which would have survived in the absence of these shocks. Regardless, this does not affect our original conclusion about the lack of significant difference in survival (or destruction) between HD and MHD simulations, hence, we leave the investigation for causes of this discrepancy to future studies. 

Another property which can have differences due to the inclusion of magnetic fields is the growth (destruction) rates of the cold gas. For that, we repeat the process of calculating the cold gas mass for each snapshot to obtain the temporal evolution of cold gas mass for all the different simulations with $\mathcal{M} = 0.5$, and Fig.~\ref{fig:cold_gas} shows the same. We find a lack of drastic differences in the growth rates of the simulations that are well within the survival regime. We see slight differences (within a factor of 2) in simulations close to the transition regime, but it is still less than the order of magnitude differences seen in and expected from the TRML simulations. We also plot the expected growth curve for the surviving cases, using $t_{\rm grow}$ from equation~(7) in \citet{Gronke2022SurvivalMedium} and the mass growth equation for ``fragmented'' growth, $M_{\rm cold} = M_{\rm cold,0}~e^{(t/t_{\rm grow})}$, as

\begin{align}
    t_{\rm grow} \equiv \alpha \chi \mathcal{M}_\mathrm{s, hot}^{-1/2} \bigg( \frac{R_\mathrm{cloud}}{l_\mathrm{shatter}} \bigg)^{1/2} \bigg( \frac{R_\mathrm{cloud}}{L_\mathrm{box}} \bigg)^{-1/6} \tcoolcold
\end{align}
\noindent where, $\alpha = 0.5$ is a fudge factor.

This lack of significant difference in the cold gas mass evolution and survival criteria between turbulent box simulations with (MHD) and without (HD) magnetic fields is surprising, in view of the results from TRML simulations in the previous (cf. \S\ref{sec:KH_results}), where we found that the mixing rates are (highly) suppressed when magnetic fields are introduced into the same simulations. This dichotomy in the results can be confusing, and we discuss a possible solution to this in the discussion section (cf. \S\ref{sec:disc_mass_transfer}).

\subsection{Cold gas distribution and morphology}
\label{sec:cold_morph}

In the previous subsection, we showed how the presence of magnetic fields seems to have only a marginal effect on the cold gas mass growth and survival. Still, the magnetic fields are not entirely inconsequential. The magnetic fields can affect the gas flow and vice-versa due to effects like flux-freezing. We also saw in the above section that the morphology of the cold clouds is different between the simulations with (MHD) and without (HD) magnetic fields (cf. Fig.~\ref{fig:turb_snapshot}). In this section, we present such differences and quantify these differences in the morphology and distribution of the cold gas.

\begin{figure}
	\includegraphics[width=\linewidth]{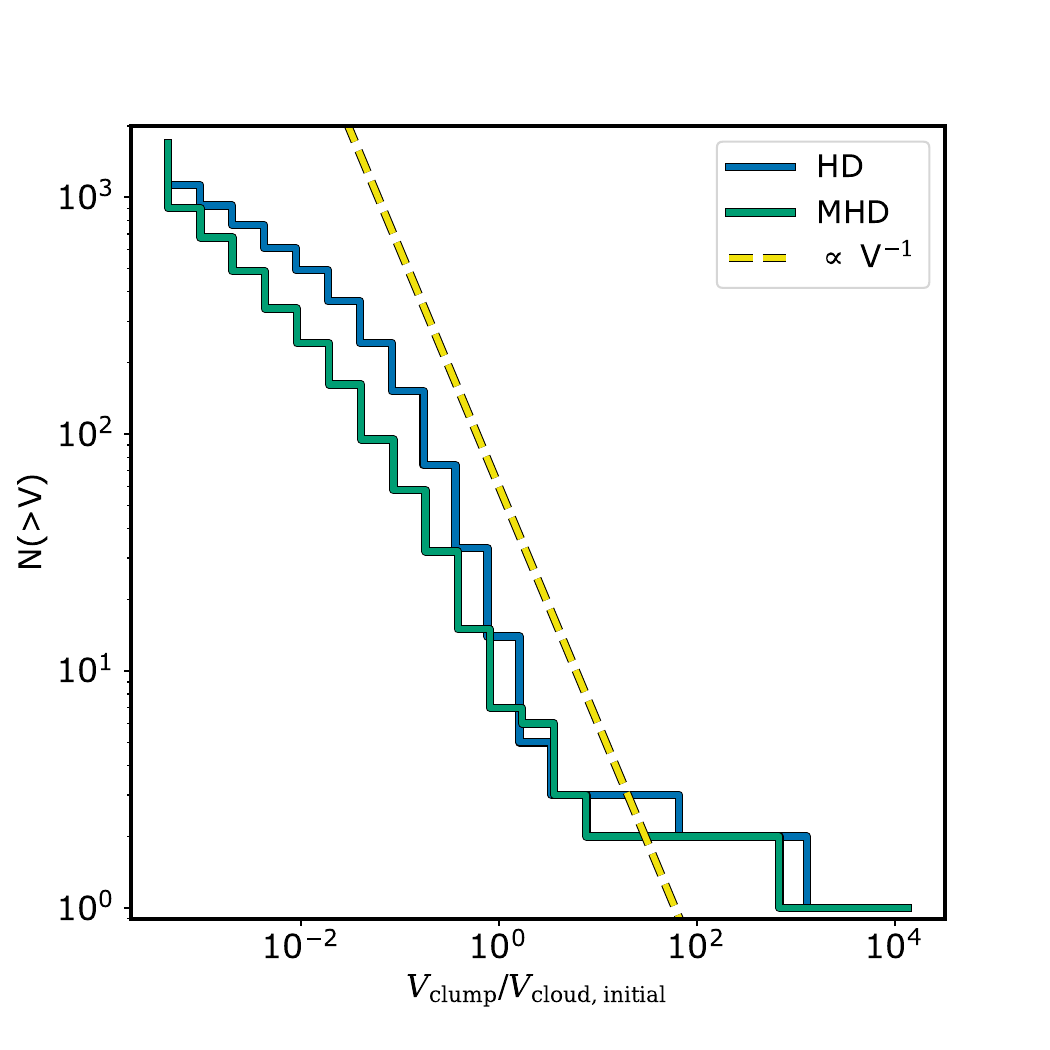}
    \caption{Cumulative number distribution for HD-MHD simulation pair with $\mathcal{M} = 0.5$ and $R_{\rm cloud} = 310 l_{\rm shatter}$. This shows the marginal difference in the overall distribution of clump sizes, and also that the distribution matches the distribution of $\propto V^{-1}$, found in previous studies.}
    \label{fig:cold_clump_hist}
\end{figure}

\begin{figure}
	\includegraphics[width=\linewidth]{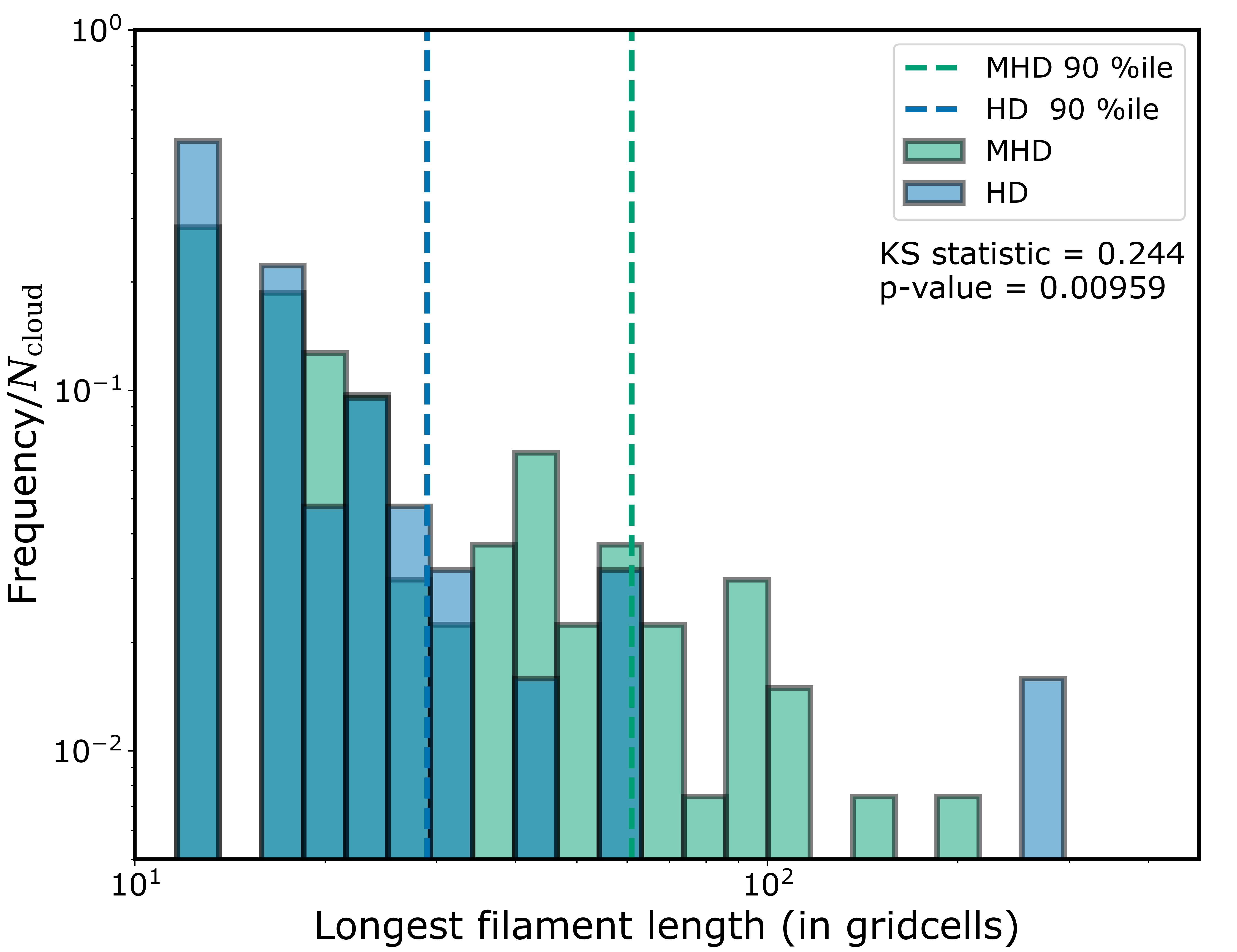}
    \caption{Histogram of longest shortest distance in the neighbourhood graph of every clump in a snapshot from the turbulent box at $\mathcal{M}_{\rm turb}=0.5$, with and without magnetic fields. This figure gives a lower limit on the difference in the filamentariness of the cold gas clumps in the two cases. We find at cold gas clumps can get more filamentary in the presence of magnetic fields, by about a factor of 2.}
    \label{fig:clump_size}
\end{figure}

\textchange{Turbulent transport has been a long-standing field of research in fluid dynamics. In a turbulent medium, the stochastic motions can transport, break, coalesce or mix the cold gas clouds. This results in a wide variety of cold gas cloud morphology. \citet{Gronke2022SurvivalMedium} calculated the cloud size distribution in a hydrodynamically turbulent medium and found a power law, $N(>V) \propto V^{-1}$ (which has also been found in larger scale simulations, e.g., \citealp{Tan2023CloudWinds}). As we saw a significant difference in the cold gas structure in Fig.~\ref{fig:turb_snapshot}, we check if the visual difference in cold gas morphology between HD-MHD simulations is reflected in cold gas size distributions.}

We calculate the cumulative number distribution of the cold clumps in a set of HD-MHD simulations with $\mathcal{M}_\mathrm{s, hot} = 0.5$ and $R_{\rm cloud} = 310 l_{\rm shatter}$. Similar to \textchange{Fig.~\ref{fig:cold_gas}}, we define cold gas as the gas with temperature below $ 2T_{\rm floor} = 8 \times 10^4$ K and use feature labelling functionality in SciPy's \citep{Virtanen2020SciPyPython} \texttt{ndimage} as the clump finding algorithm to identify the cold clumps. We determine the volume of the obtained clumps and use it to calculate the cumulative number distribution, shown in Fig.~\ref{fig:cold_clump_hist}. We find that the distribution is well approximated by a power-law with slope -1, \textchange{i.e. $N(>V) \propto V^{-1}$,} while deviating at the two extremes of the volume range due to resolution limits \textchange{ at lower volumes and statistical under-representation at higher volumes.} This matches the results from \citet{Gronke2022SurvivalMedium}. We also see that the number distribution does not show a drastic difference between the HD and MHD simulations, apart from a slight excess of small clumps in the HD simulation. \textchange{This means the cold gas clumps in MHD simulation are not significantly smaller or larger in volume than its HD counterpart.}

Even though the difference in number distribution is minor, Fig.~\ref{fig:turb_snapshot} clearly shows significant morphological differences between the cold gas structure in HD and MHD simulations. Visually, the cold gas in  MHD simulations has a much more filamentary shape, while it has a more clumpy cold gas morphology in the corresponding HD simulations. We quantify this filamentariness of the cold gas structures as the length of the longest ``shortest path'' possible within the clump. 

\begin{figure*}
	\includegraphics[width=\textwidth]{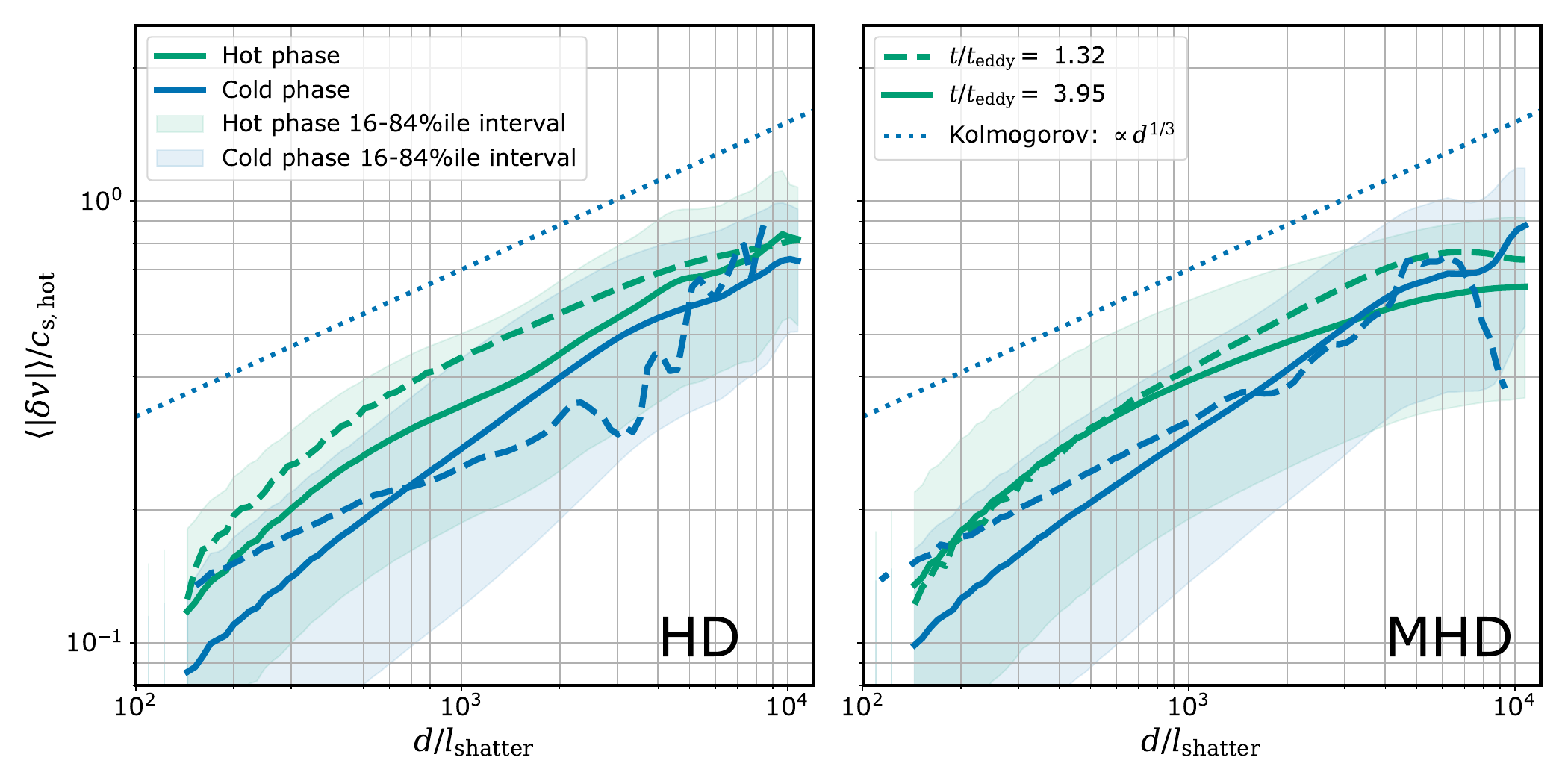}
    \caption{Velocity structure function (VSF) for hot (green line) and cold (blue line) gas phases in a set of simulations with (MHD) and without (HD) magnetic fields, at $\mathcal{M} = 0.5$ and $R_{\rm cl} = 310 l_{\rm shatter}$. The dashed and solid lines show the VSF at different times, $t = 1.32 t_{\rm eddy}$ and $3.95 t_{\rm eddy}$ after introducing the cold gas cloud. This shows the decreasing difference in VSF of the two phases with time, in both cases, which means that the two phases are kinematically well-connected. We also find a smaller early-time difference between the hot and cold gas VSF for the MHD simulation, indicating a better kinematic connection in that case.}
    \label{fig:vsf}
\end{figure*}

To do so, we first identify the individual clouds, as done for Fig.~\ref{fig:cold_clump_hist}, and create a ``neighbourhood graph'' for each clump using an adjacency matrix. In this ``neighbourhood graph'', each gridcell inside the cloud is a node and two nodes are connected with an edge if the two share a face. We calculate the shortest path between every node in this neighbourhood graph, and take the longest from this list of shortest paths as the required longest ``shortest path''. As many of the largest clumps contain $\gtrsim~50,000$ gridcells, we have to use a faster way which can give a close enough answer instead of using the brute force method. The slowest step of the method is the shortest path calculation among each pair of nodes. Hence, to speed up this step, we only consider every 4$^\mathrm{th}$ node to identify the two points with the longest ``shortest path", and later recalculate the path using the full graph with all nodes to get the final length. This optimised method drastically reduces the number of path calculations and makes this analysis computationally feasible. We also find negligible differences between the full brute force method and the optimised method in our tests. This is because the points with the actual longest ``shortest path'' is usually in close vicinity, likely within 4 grid lengths, unless the reduced graph is drastically different, which is rare.

We repeat this process for each clump in the MHD and HD simulations with $\mathcal{M}=0.5$ and $R_{\rm cloud} = 310~l_{\rm shatter}$ and plot the histogram of the obtained longest ``shortest path'' in Fig.~\ref{fig:clump_size}. We find that the 90\%ile of this length distribution for MHD simulations is longer by a factor of 2 compared to the corresponding HD simulation. Assuming, a constant volume of a cylindrical cold gas clump, which is reasonable as shown in Fig.~\ref{fig:cold_clump_hist}, an increase of 2 in length corresponds to an increase of $\approx 2.8$ in the length-to-width ratio of the clump. This method under-quantifies the filamentariness of the clumps, as the connected filamentary structures that are shorter than the main filament are not included. A full tree-based filament analysis will be the ideal method for this analysis, but we leave the detailed study of filamentariness to future investigations.

We find that the p-value for the two length distributions in HD and MHD is lower than 0.05, which means we can consider the filamentariness of HD and MHD simulations to have different underlying distributions. The KS statistic quoted in Fig.~\ref{fig:clump_size} quantifies the difference between the two distributions and is linked to the p-value. The higher the KS statistic, the lower the p-value. As our number of samples is limited by the number of clouds in the simulation, we will need a bigger box and longer runtime to improve the confidence level of this conclusion. 

\textchange{So, we conclude that, even though the cold gas clouds in the MHD simulation are similar in volume and its distribution, compared to their HD counterparts, they are significantly more filamentary in their morphology.}

\begin{figure*}
	\includegraphics[width=\textwidth]{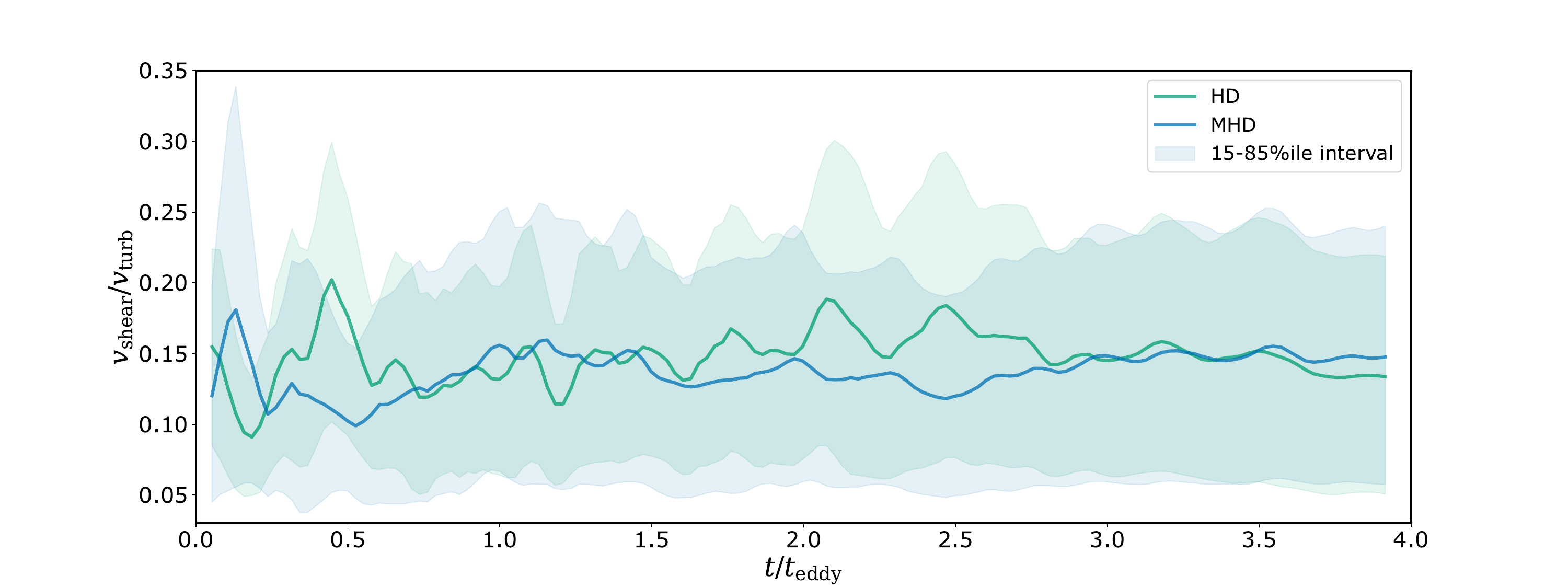}
    \caption{Evolution of average shear at clump boundaries in a set of HD and MHD simulations with $\mathcal{M} = 0.5$ and $R_{\rm cl} = 310 l_{\rm shatter}$ (same as Fig.~\ref{fig:vsf}). The shaded regions show the corresponding 15-85\%ile intervals. The figure also shows the shear on the clump boundaries is about an order of magnitude lower than the turbulence velocity in the simulations. Also, on average, clumps in the MHD simulation seem to have a marginally lower, but very similar shear, in comparison to HD simulations. }
    \label{fig:shear}
\end{figure*}

\subsection{Cold gas entrainment}
\label{sec:cold_vsf}

In a multiphase environment, the motion of one phase can affect the motion of another via drag forces or mixing-induced momentum transfer \citep[see, e.g.,][]{Gronke2020HowWind,2021ApJ...911...68T}. This means the phases can be kinematically linked. In addition to that, due to flux freezing in the gas, the magnetic fields can increase the extent of this kinematic link. A good way to check for this is the first-order velocity structure function (VSF). It quantifies the average difference in velocities of gridcells separated by a given distance. The difference in the VSF of the two phases corresponds to a lack of link in the kinematics between the two, while a smaller difference corresponds to a greater kinematic link.

We calculate the velocity structure function (VSF) of the hot and cold phases of gas in simulations with (MHD) and without (HD) magnetic fields at different times. First, we calculate pairwise distances and velocity difference magnitudes between each pair of gridcells. Again, due to computational constraints, we cannot use all cells for the pairwise calculations, so we randomly choose $2\times10^4$ gridcells for this calculation. Then, the list of pairwise velocity differences is binned according to the pairwise distances and we plot the average of the velocity differences in each bin as the VSF in Fig.~\ref{fig:vsf}. In general, we find a higher value of VSF in the hot gas than the cold gas for both the simulations at all different times in the evolution, as seen in other idealised simulations \citep{Gronke2022SurvivalMedium, Mohapatra2022VelocityClusters},  and even some observations \citep{Li2022TurbulenceGalaxy}. 

\textchange{For structure function calculations with steep slopes, \citet{2023MNRAS.518..919S} had found a 2-point stencil to be unconverged and suggested the use of higher-order stencils. But, the slope of VSF for Kolmogorov turbulence (1/3) is shallow enough for a 2-point stencil to be converged. Hence, we use a 2-point stencil for all our VSF calculations.}

This difference between the hot and cold VSF is much larger in HD simulation, on the left panel of Fig.~\ref{fig:vsf}, while in MHD simulations (right panel of Fig.\ref{fig:vsf}) the difference is much more subtle with both hot and cold VSF comfortably within 16-84 \%ile range of each other. This shows that the cold phase is, in general, better entrained in MHD simulations compared to HD simulations. This is likely due to the flux-freezing of the magnetic fields that can result in a more efficient kinematic connection between the hot and cold phases, as mentioned before. \textchange{In presence of flux-frozen magnetic fields, any relative motion between the two phases encounters an enhanced drag force \citep{McCourt2015MagnetizedWind}}. We also find that, even though the VSF of the different phases start off differently in HD-MHD simulations, they end up with very similar hot and cold VSF profiles in both cases. This means, given enough time, both simulations reach a similar extent of entrainment. Still, we do note that the entrainment is faster in MHD simulations, compared to HD, as shown by the faster decreases in the difference between hot and cold medium VSF in MHD simulations. This result indicates high entrainment of cold gas in hot gas, albeit an imperfect one. More importantly, it also points to a more efficient and faster entrainment of the cold gas in the presence of magnetic fields, with both HD and MHD simulations reaching an equivalent entrainment state, given enough time.

Further, we calculate the average shear at the cold gas clump boundaries for each snapshot in the HD and MHD simulations with $\mathcal{M} = 0.5$ and $R_{\rm cl} = 310 l_{\rm shatter}$ using \texttt{yt}\citep{Turk2011Yt:DATA}. Fig.~\ref{fig:shear} shows the evolution of this average and the 15-85\%ile interval of the shear at clump boundaries with time. We find that the shear at the clump boundaries is, in general, about one order of magnitude lower than the turbulent velocity in the simulation. This again indicates a high entrainment of the cold gas. Also, the slightly lower values of average and 85\%ile value of shear for MHD simulations suggest a more efficient and faster entrainment in the presence of magnetic fields.

\subsection{Magnetic fields strength and structure}
\label{sec:cold_B_field}

In MHD simulations, magnetic fields are kinematically very important, as the gas flows affect the magnetic fields and vice-versa. Apart from affecting the kinematics, magnetic field structure can also affect other processes like thermal conduction and cosmic ray transport in an astrophysical media \citep[e.g.,][]{Kempski2023CosmicReversals,RuszkowskiPfrommerCRREVIEW}.

\textchange{The turbulent motions can result in a local dynamo effect, leading to amplification of magnetic fields in MHD simulations. The extent of this amplification can vary in the different phases due to differences in the Alfv\'enic wave speed ($ v_{\rm A} = B/\sqrt{\rho}$) and turbulent velocities. On top of the dynamo effect, the compression of gas during its cooling can also cause amplification during cold gas formation, due to flux-freezing.}

\begin{figure*}
	\includegraphics[width=\textwidth]{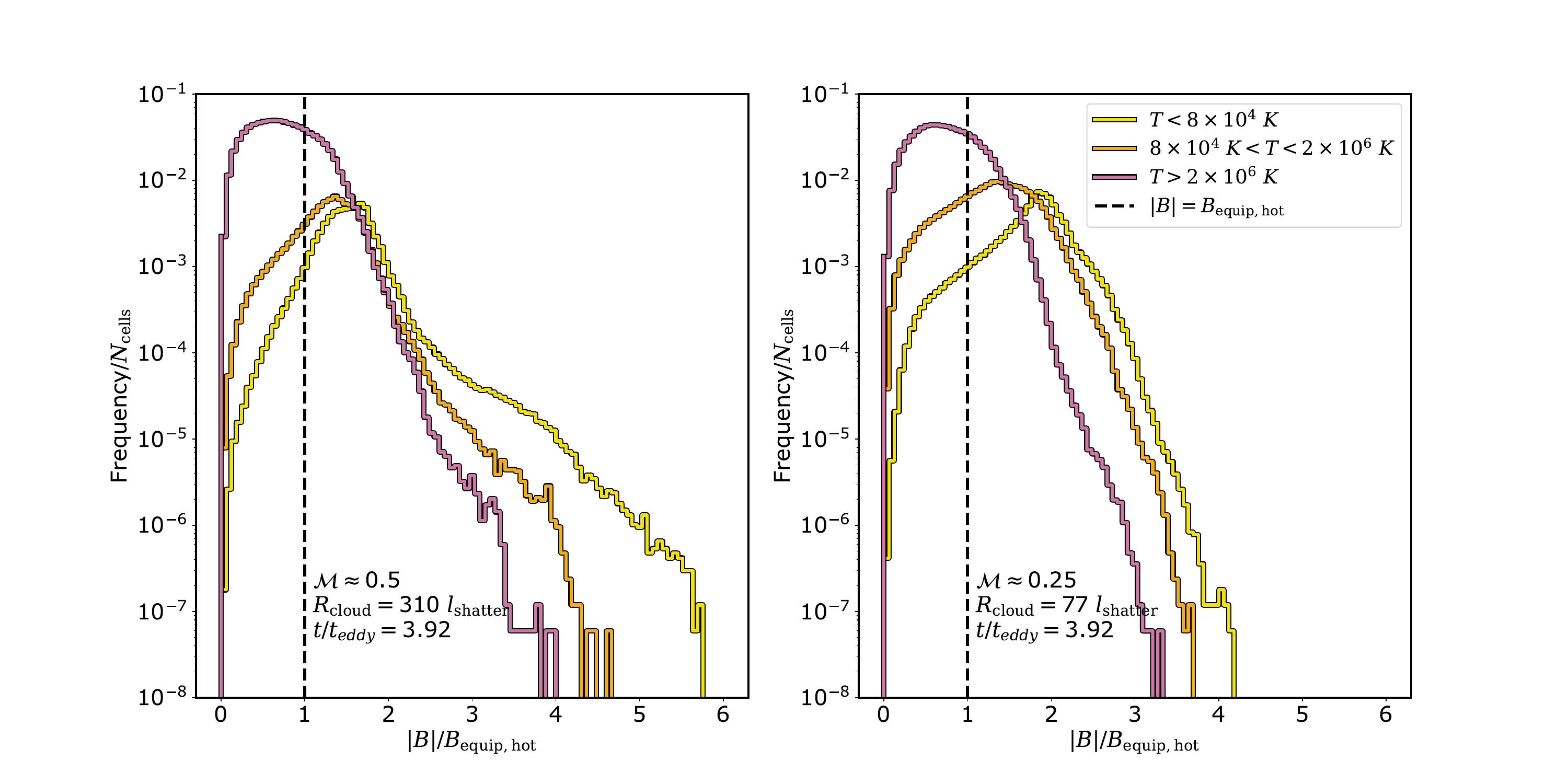}
    \caption{Histogram of magnetic field strength in gas within different temperature ranges, namely hot ($T>2 \times 10^6$ K), mixed ($8 \times 10^4~\mathrm{K}<T<2 \times 10^6~\mathrm{K}$), and cold ($T<8 \times 10^4$ K) gas, for two simulations where the cloud gas cloud survives, $t=3.92 t_{\rm eddy}$ after its introduction. \textbf{Left} $\mathcal{M} = 0.5$, $R_{\rm cl} = 310 l_{\rm shatter}$. \textbf{Right} $\mathcal{M} = 0.25$, $R_{\rm cl} = 77 l_{\rm shatter}$. The dashed vertical line corresponds to the equipartition magnetic field strength, achieved in the hot ambient gas at the end of driving the turbulence. This shows that the magnetic fields are significantly amplified as the gas cools down to a lower temperature. We discuss the possible causes of this amplification in \S~\ref{sec:disc_Bfield}.}
    \label{fig:Bfield_hist}
\end{figure*}

\textchange{To examine these differences,} we check the distribution of magnetic field strengths in the different phases. We define the cold phase as the gas with temperature $T<2T_{\rm floor}=8\times 10^4$ K, hot phase as $T> 0.5 T_{\rm amb} =2\times 10^6$ K, and mixed gas as the gas with temperatures in between, i.e. $8\times 10^4~\mathrm{K} < T < 2\times 10^6~\mathrm{K}$. Fig.~\ref{fig:Bfield_hist} shows the distribution of magnetic field strengths in these three gas phases, for simulations of three turbulent Mach numbers and with two cloud radii. The exact distribution has a non-trivial dependence on factors like the turbulent Mach numbers and cold gas growth rate. But in all cases, the mixed and cold gas magnetic strength distributions are centred at stronger magnetic fields, while the hot gas magnetic strength is centred around weaker magnetic fields, with the dashed line showing the equipartition magnetic field strength. This higher magnetic field strength in cold and mixed gases could be due to three possible processes: turbulent local dynamo in the dense gas as the equipartition magnetic field is higher for a denser gas moving at similar velocities, flux-freezing accompanied by compression due to cooling, and magnetic draping caused by the relative motion between the dense gas and magnetic fields. We discuss more about these processes, and possible order of importance in the discussion section (c.f. \S\ref{sec:disc}).

\begin{figure}
	\includegraphics[width=\columnwidth]{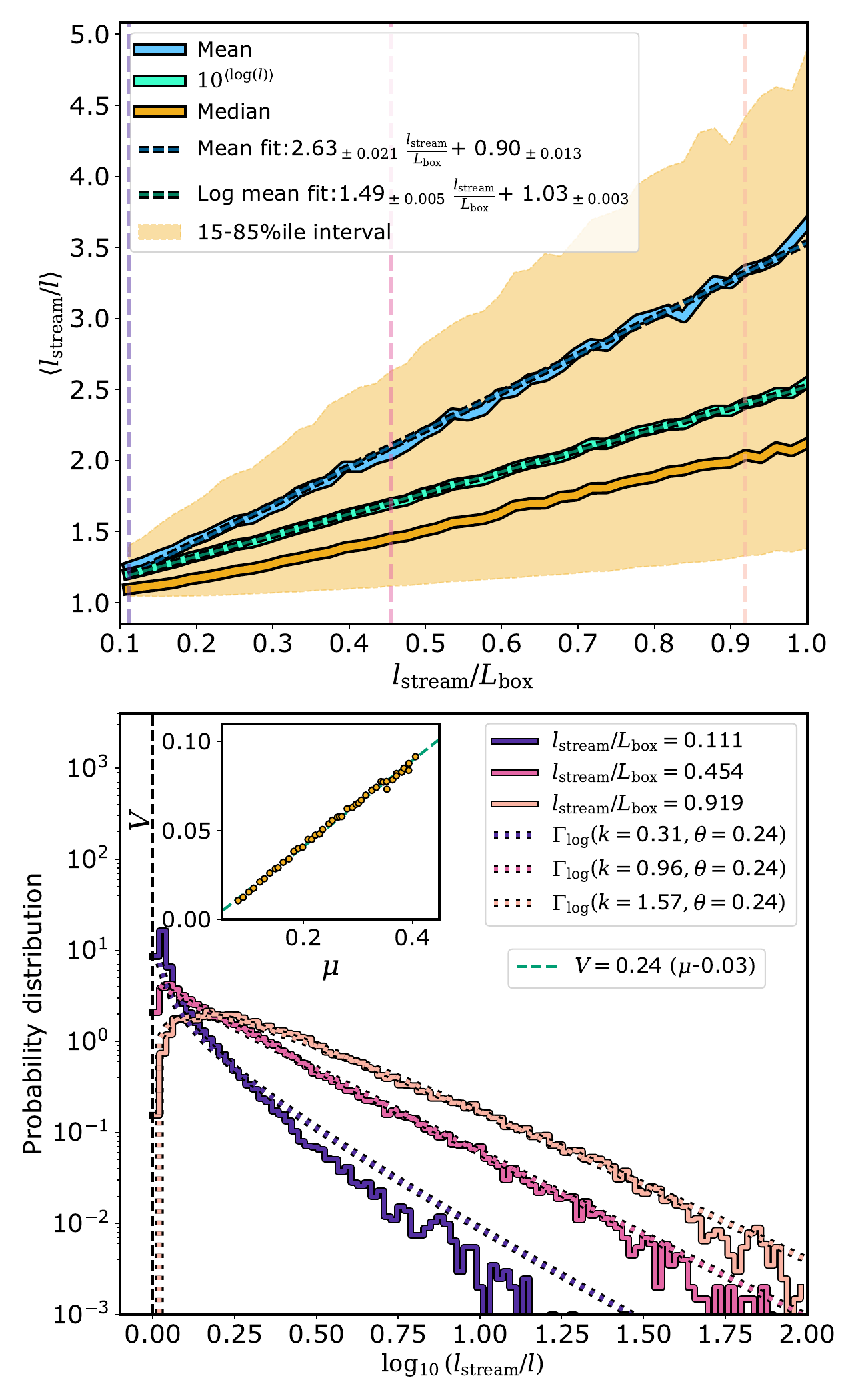}
    \caption{\textbf{Top} Average, median and $10^{\text{average of logarithm of}}$ entanglement, i.e. $l_{\rm stream}/l$ for different streamline lengths ($l_{\rm stream}$). The dashed lines show the corresponding best linear fits and the shaded region shows the 15-85\%ile interval. The general trend of increasing entanglement for longer and longer streamline lengths indicate a fractal-like structure of the magnetic field lines, discussed further in \S~\ref{sec:disc_Bfield}. \textbf{Bottom inset} Points denote the mean and variance of $\log_{10}{(l_{\rm stream}/l)}$ and the green dashed line shows the linear fit, $V = 0.24 (\mu - 0.03)$. We use this relation to calculate the shown probability distribution. \textbf{Bottom} Solid lines show the probability distributions of different values of entanglement, $\log_{10}{(l_{\rm stream}/l)}$, for three values of streamline lengths. The dashed lines show the corresponding calculated $\Gamma$ distributions, with the parameters mentioned in the legend. This shows the close agreement between the estimated and calculated probability distributions. There are some deviations for the probability distribution of small streamline length, which is discussed further in \S~\ref{sec:cold_B_field}.}
    \label{fig:Bfield_tangle}
\end{figure}

\textchange{Flux-freezing and the turbulent motions can result in a very tangled magnetic field structure. These tangled mangetic fields can have many consequences including reduced thermal conduction and slower cosmic ray transport. Presence of multiphase gas in a turbulent medium can add further complexity to the magnetic morphology, due the magnetic field strength distributions in different phases, as we show earlier in this section. To better understand this, we study the structure of the magnetic fields using magnetic field streamlines.}
We use \texttt{yt} \citep{Turk2011Yt:DATA} to calculate 10000 streamlines for 100 different streamline lengths ($l_{\rm stream}$) between $0.01 - 1 L_{\rm box}$. We calculate the displacement between the two endpoints of the streamlines ($l$), to get the ratio  $l_{\rm stream}/l$. This ratio denotes the extent of entanglement of the magnetic field. A $l_{\rm stream}/l=1$ indicates a perfectly untangled streamline, with higher values denoting a higher extent of entanglement. We repeat this process for different $l_{\rm stream}/L_{\rm box}$, and obtain distributions of the extent of entanglement ($l_{\rm stream}/l$) for each $l_{\rm stream}/L_{\rm box}$. We calculate the mean, median and mean of the logarithmic lengths in each $l_{\rm stream}/l$ distribution.

The upper panel of Fig.~\ref{fig:Bfield_tangle} shows the trend of mean, median, mean(log) and 15-85\%ile interval for each $l_{\rm stream}/L_{\rm box}$ value. We find that all the metrics of ensemble value of the ratio $l_{\rm stream}/l$ increase linearly with $l_{\rm stream}/L_{\rm box}$. The upper panel of Fig.~\ref{fig:Bfield_tangle} shows the close approximation of the linear trend for mean and mean(log). This means the extent of the entanglement increases linearly with the length of the streamline. This property could be a sign of fractal-like behaviour of the field lines down to a certain threshold at small scales. We discuss this further in the discussion section (cf. \S\ref{sec:disc}). Note that the asymmetrically located $15$th and $85$th percentiles with respect to the median indicate a long tail towards longer $l_{\rm stream} / l$.

To show this more explicitly, we further choose three different streamline lengths, shown as vertical dotted lines in the upper panel of Fig.~\ref{fig:Bfield_tangle}, and recalculate the streamlines for 10$\times$ more ($10^5$) starting points. We repeat the above mentioned process to  calculate the $l_{\rm stream}/l$ values and calculate the probability distribution function of the entanglement, $l_{\rm stream}/l$, for these three values. We plot these probability distribution functions as solid lines in the bottom panel of Fig.~\ref{fig:Bfield_tangle}. Even though we find some minor deviations at higher entanglement values due to insufficient counts caused by the reduced number of samples, the histogram at lower entanglement values is robust and fairly well converged with the number of streamlines. 

Next, we attempt to find an analytic form for the different probability distribution functions (PDF) that we found earlier for the  $l_{\rm stream}/l$, in the bottom panel of Fig.~\ref{fig:Bfield_tangle}. We calculate the variance ($V$) and mean ($\mu$) of $\log_{10}{(l_{\rm stream}/l)}$ and find a strong linear relation between the two as $V = 0.242(\mu - 0.031)$. For a $\Gamma$ distribution, the variance and mean are given by $V=k\theta^2$ and $\mu=k\theta$, where $k$ and $\theta$ are the shape and scale parameters, respectively. This means, our $V-\mu$ relation for $\log_{10}{(l_{\rm stream}/l)}$ matches the properties of a $\Gamma$ distribution with $\theta = 0.242$, and $k = \mu/\theta = \mu/0.242$. Using the linear fit for the $\mu$, shown in upper panel of Fig.~\ref{fig:Bfield_tangle}, and the equation of the $\Gamma$ distribution, a fit for the PDF of $l_{\rm stream}/l$ for a given $l_{\rm stream}/L_{\rm box}$ is a $\Gamma$ distribution for $x = \log_{10}{(l_{\rm stream}/l)}-0.031$ with $k \approx \log_{10}(1.5 l_{\rm stream}/L_{\rm box} + 1)$ and the $\theta=0.242$ mentioned above.


The bottom panel of Fig.~\ref{fig:Bfield_tangle} shows the analytical form with the dotted lines. We find that the analytical form agrees very well with the PDF for long streamlines. But for shorter streamlines, at intermediate values of entanglement ($l_{\rm stream}/l$), it overestimates the PDF at intermediate entanglement values in the tail. This can be an indication towards a different underlying analytic form for PDF, which is equivalent to the $\Gamma$ distribution at longer streamlines. Or, it can be due to resolution effects as they start to become more important for short highly entangled streamlines. We leave a deeper investigation of this for future studies.

\textchange{As the charged particles tend to gyrate around and follow the magnetic field lines, analytic form for the magnetic field morphology, similar to the ones we find above, can be used in developement of models for their transport through a multiphase turbulent medium.}

\subsection{Synthetic absorption line spectra}
\label{sec:mock_spec}

As shown in the results section of turbulent boxes (cf. \S\ref{sec:turb_results}), we know that the morphology and details of kinematics can differ significantly between the simulations with (MHD) and without (HD) simulations. This difference can affect observational probes like predicted quasar absorption line spectra, because the column density and Doppler shift, the two major features of lines, can be affected by these differences in morphology and kinematics. We investigate these effects and their observational consequences in this section. 

\begin{figure}
	\includegraphics[width=\columnwidth]{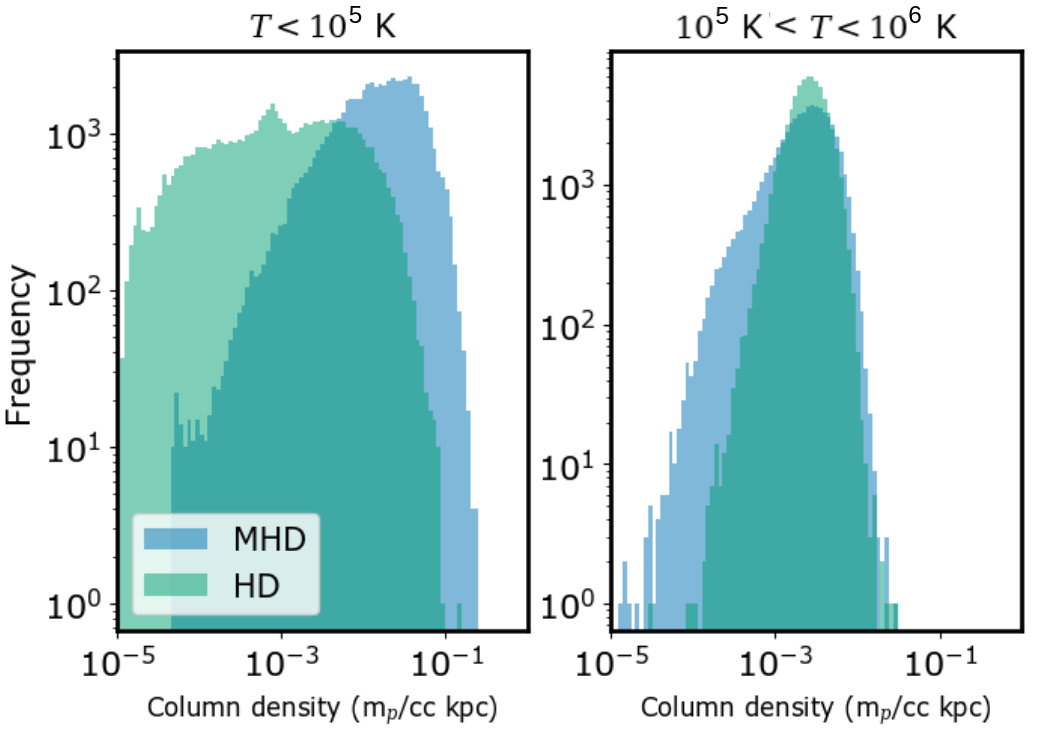}
    \caption{Column density distribution of cold ($T<10^5$ K, left panel) and intermediate ($10^5~\mathrm{K}<T<10^6~\mathrm{K}$, right panel) temperature gas in HD (in green) and MHD (in blue) simulations, with $\mathcal{M} = 0.5$ and $R_{\rm cl} = 310 l_{\rm shatter}$. This shows that the column densities for the above cases are within the observationally expected column densities for absorption spectra in a circumgalactic environment. It also shows that the lower end of column density distribution for cold temperature gas has a higher extent of difference between the HD and MHD simulations. This makes an absorption line tracing the cold gas a prime candidate for looking at observational differences between the HD and MHD simulations.} 
    \label{fig:column_density}
\end{figure}

Fig.~\ref{fig:column_density} shows the distribution of column densities of cold ($T<10^5$ K) and intermediate/mixed ($10^5<T<10^6$ K) along one of the dimensions of the box for a set of HD-MHD simulation with $\mathcal{M}_\mathrm{hot, turb}=0.5$ and $R_\mathrm{cloud}=310~l_\mathrm{shatter}$, same as Fig.~\ref{fig:turb_snapshot} and \ref{fig:vsf}. Note that because $l_{\rm shatter}\propto n^{-1}$ \citet{McCourt2018AGas} the column densities simulated can be directly compared to observations. We still find that the column density histogram of the cold gas shows a greater extent of difference between the set of HD and MHD simulations, compared to the intermediate gas. We also see that most of the difference shows up at lower values of column density. So, we expect to see some difference between the HD and MHD simulations in observational probes of cold gas, for example, LOS absorption due to MgII, at lower equivalent widths.

In order to investigate the observational consequences of the differences between HD and MHD simulations, we create mock LOS absorption spectra using \texttt{Trident} \citep{Hummels2017Trident:Simulations} on the same set of HD-MHD simulations as Fig.~\ref{fig:turb_snapshot}, \ref{fig:clump_size}, \ref{fig:vsf}, and \ref{fig:column_density} ($\mathcal{M}_\mathrm{hot, turb}=0.5, R_\mathrm{cloud}=310~l_\mathrm{shatter}$). Note that both of these snapshots have similar cold gas volume filling fractions.

We sample $\sim10000$ line-of-sight (LOS) spectra along one of the axes of the computational domain on a $100\times100$ grid. Due to the isotropic nature of the system, the particular choice of the axis should not affect the statistical inferences.  We use a $\Delta\lambda = 0.1$\r{A} (corresponding to a spectral resolution $R\equiv \lambda/\Delta\lambda\approx 28,000$) to create the mock absorption features for MgII at uniform solar metallicity. We select spectra which have a maximum absorbed flux of more than 0.1, over a continuum flux of 1.0, so that we are only considering LOS that pass through significant amounts of cold gas. We also exclude the spectra which have saturated features (with a flux less than 0.1) because they correspond to unnaturally large cold gas volume filling fractions, thus, leaving the number of components ill-defined. Next, we calculate the equivalent width (EW) for the MgII line complex and the number of absorption features for each LOS, as the area under the continuum and the number of minima in a spectrum, respectively. Fig.~\ref{fig:los_churchill} shows the 2D distribution of EW and the number of absorption features for the HD-MHD simulations. We also show the fit obtained for the same quantities from observations of MgII absorbers in quasar spectra in \citet{Churchill2020MgModels}. Interestingly, the more frequent regions of both the 2D histograms in Fig.~\ref{fig:los_churchill} are close to the relation found in \citet{Churchill2020MgModels}. We also find that, for the same EW, the MgII spectra of the MHD simulation tend to have a slightly higher number of absorption features, compared to the HD simulation, but these differences are marginal.


We repeated this exercise with 10x better spectrum resolution, which is much higher than that of the observed spectra, and we also included an additional CIV 1551\r{A} line. The corresponding distributions, analogous to Fig.~\ref{fig:los_churchill}, are shown Fig.~\ref{fig:MgII_highres} and \ref{fig:CIV_highres} in Appendix~\ref{sec:app_mock_spectra}). In both cases, the distributions change, but still roughly follow the observed curve.

\section{Discussion}
\label{sec:disc}

\subsection{Mass transfer rates in a magnetized, turbulent medium}
\label{sec:disc_mass_transfer}

The two results for mixing layers and turbulent box simulations shown in Sec.~\ref{sec:trml} and Sec.~\ref{sec:turb_results}, respectively, present a dichotomy. On one hand, TRML simulations show a significant suppression in the mixing of two phases, and on the other hand, the turbulent box simulations show the lack of a similar difference in mixing rate, as shown by the cold gas growth rates and survival criterion.

To resolve this, we need the answer to the question, \textit{what is the primary mechanism of mixing of two phases?} The mixing happens when the multiphase structures get small enough to reach the ``dissipation scale'' where the molecular diffusion is fast enough to mix the two phases  \citep[Obukohov-Corrsin phenomenology;][]{Oboukhov1949StructureFlows, Corrsin1951OnTurbulence}. There can be multiple ways to reach such small scales, and one of these is via vorticity. Vorticity or vortices can stretch, fold and transport, in other words, it ``stirs'' and stretches the structures, eventually reaching the small scales where molecular diffusion can take over and ``mix'' the two phases \citep{Villermaux2019MixingStirring}. In theory, this vorticity does not have to be part of turbulence, but in the high Reynold's number of fluids, as is the case in astrophysical mediums, the vortices generally become turbulent. This causes a faster stirring of the multiphase structures, and a more rapid increase in the surface area and decrease in structure width, resulting in more efficient diffusion. Hence, in an astrophysical medium, the main mechanism of mixing is expected to be turbulent mixing. In principle, the mixing should only depend on the turbulence and be independent of the source of turbulence.

In a Kelvin-Helmholtz or TRML setup, the initially structured vortices quickly give way to a turbulent mixing layer. The turbulence in this mixing layer is the key mechanism of stirring and mixing the two phases. This lead to this chain of processes: KH instability $\rightarrow$ Turbulence $\rightarrow$ Mixing. When magnetic fields are introduced in the system, depending on their orientation, they hinder the link between the KH instability and turbulence by slowing down the rate at which the turbulence in the mixing layer is driven. But as we show above, importantly, the magnetic fields do not affect the other link that connects turbulence with mixing. So, we expect to see a tight correlation between the turbulent velocity in the mixing layer and the extent of mixing/cooling that is occurring in the mixing layer, regardless of the magnetic field orientation or strength (cf. Fig.~\ref{fig:Q_vs_vturb}). Hence, we conclude that suppression of mixing in TRML simulations in magnetic fields is due to the reduced driving of turbulence in the mixing layer, which in turn leads to a reduced mixing of the two phases. 

The situation is different in our turbulent box setup. There the system is the driven turbulence that cascades from the largest (box-size) scales to the smallest (gridcell-size) scales. This implies that $u'$ is fixed and since the mixing and cooling rate only depends on $u'$ directly, we obtain similar growth rates in the HD and MHD cases -- explaining the unaltered survival criterion and mass transfer rates found (cf. Figs.~\ref{fig:cloud_survival} and \ref{fig:cold_gas}, respectively).

We also find direct evidence that the turbulent, cascading $u'$ is responsible for the mixing, and not the (also in the turbulent box present) hydrodynamical instabilities seeding smaller-scale turbulence. Firstly, the velocity structure functions of both the hot and the cold medium follow each other closely (cf. Fig.~\ref{fig:vsf}) indicating near perfect entrainment of the cold gas. Secondly, we also show explicitly the  shear between cold and hot gas (c.f. \S\ref{sec:cold_vsf}) being small, i.e., the cold gas is well-entrained in the hot ambient gas. This means the shear is minimal, resulting in a lesser extent of turbulence in the mixing layer between the two phases. 
If solely the shear would be responsible for the mixing and cooling, we estimate the mass doubling time to be $\sim 5 t_{\rm eddy}$ for the turbulent box with $\mathcal{M}=0.5$ and $R_{\rm cl} = 310 l_{\rm shatter}$, which is about an order of magnitude longer than actually found in the simulations (using the TRML scaling relations of \citealp{Tan2021RadiativeCombustion} for each surface cell on the cold gas clump).

In summary, we find the $u'\rightarrow \dot m$ relation to be universal in HD and MHD (and consistent with high-resolution TRML studies; \citealp{Fielding2020MultiphaseLayers,Tan2021RadiativeCombustion}). However, magnetic fields prevent instabilities to form in the mixing layer setup leading to a lower $u'$ and thus to a decreased mass transfer rate. When the extent of turbulence is fixed by larger scales -- as done in the turbulent box -- the magnetic fields cannot suppress the mixing leading to comparable luminosities in the HD and MHD runs.

\begin{figure}
	\includegraphics[width=\columnwidth]{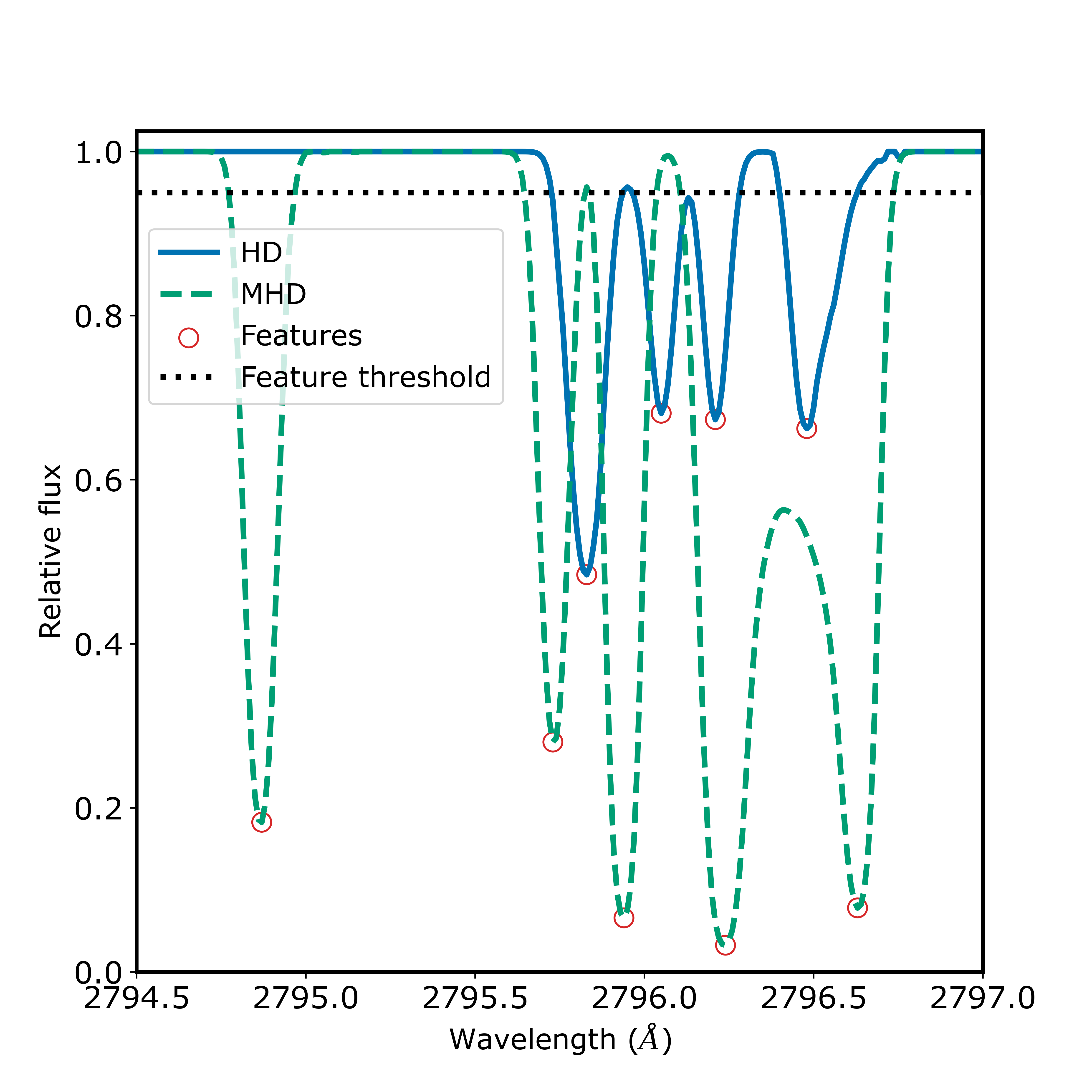}
    \caption{An example line-of-sight MgII 2796 \r{A} absorption mock spectra with $\Delta\lambda=0.01$\r{A}, from the HD (blue solid line) and MHD (green dashed line) simulations with $\mathcal{M}=0.5$ and $R_{\rm cl} = 310 l_{\rm shatter}$ (same as Fig.~\ref{fig:column_density}). The dotted black line shows the threshold of the minimum absorbed flux of a feature, and the red circles show the features that we consider for analysis. This figure is only for reference, as these are higher resolution spectra compared to the ones used in the analysis at $\Delta\lambda=0.1$\r{A}, which is closer to observational spectral resolution.}
    \label{fig:los_example}
\end{figure}
\begin{figure}
	\includegraphics[width=\columnwidth]{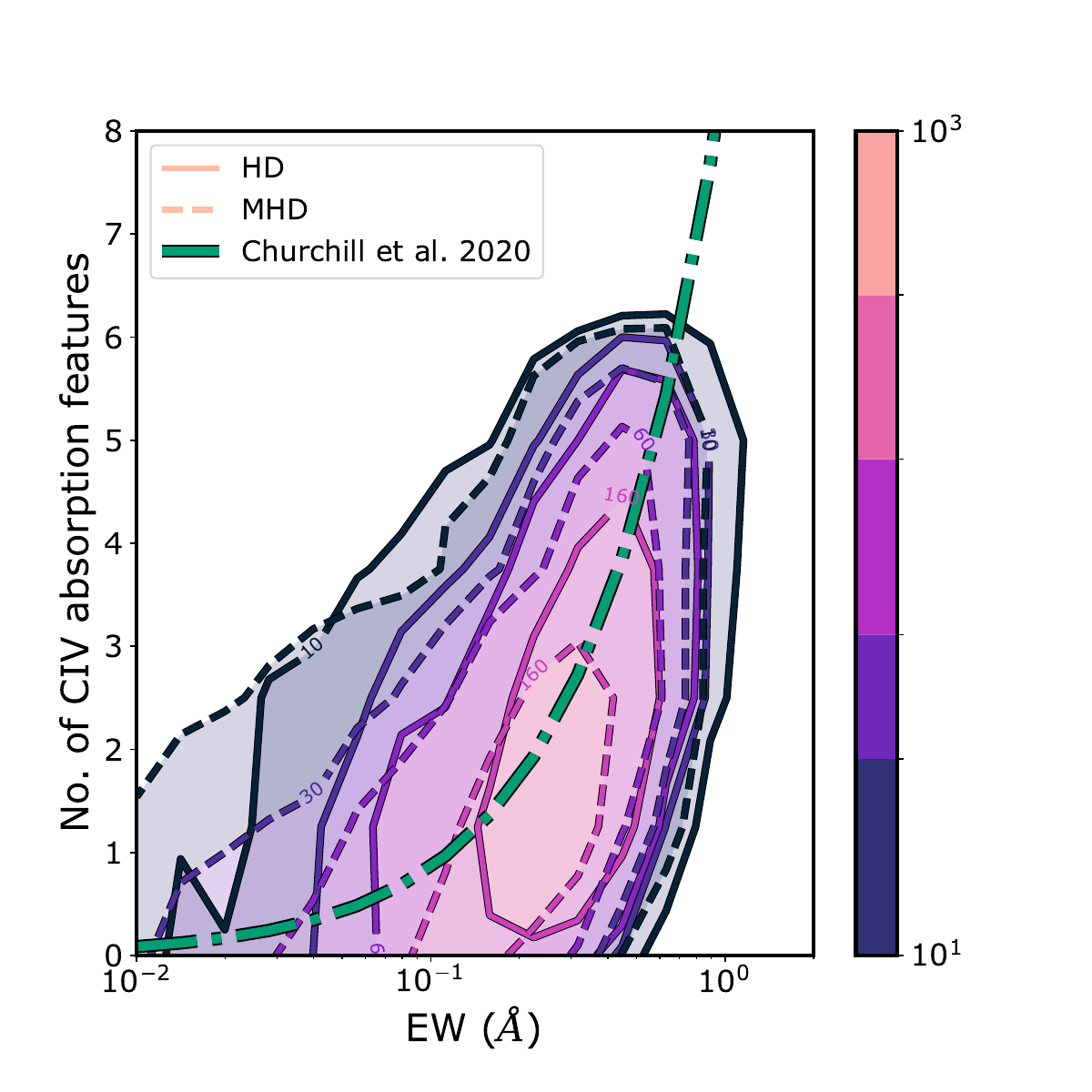}
    \caption{Contour plot of the 2D histogram of line-of-sight MgII absorption mock spectra in the number of absorption features vs. equivalent width space, for HD (solid contours) and MHD (dashed contours) simulations with $\mathcal{M}=0.5$ and $R_{\rm cl} = 310 l_{\rm shatter}$ (same as Fig.~\ref{fig:column_density} and \ref{fig:los_example}). The dash-dotted green line shows the relation found in \citet{Churchill2020MgModels}. This shows that there are only marginal differences in the overall distributions of HD and MHD simulations, despite the differences in Fig.~\ref{fig:column_density}. We also find that they agree quite well with the observed relations from \citet{Churchill2020MgModels}.}
    \label{fig:los_churchill}
\end{figure}

In realistic, astrophysical multiphase systems such as the ICM, CGM or ISM turbulence is also seeded on larger scales, then cascading downwards. In the ICM, for instance, AGN feedback is believed to play a dominant role in the stirring process leaving a characteristic imprint on the VSF \citep{Li2022TurbulenceGalaxy}. Similarly, for the CGM where both (AGN and stellar) feedback processes as well as cosmological inflow act on $\sim 100$\, kpc scales `stirring' the CGM \citep{Chen2023EmpiricalMeasurements}. The alternative `shearing layer' picture might occur in multiphase systems where bulk flows are dominant such as galactic winds and cold streams; however, since also there non-negligible turbulent is present which mixing channel is dominant is still unclear \citep{Schneider2020TheOutflows,Tan2023CloudWinds,2023MNRAS.522.1843R}.

\subsection{Magnetic field amplification and morphology}
\label{sec:disc_Bfield}

We find that the magnetic field strengths in cold and mixed gas of our MHD turbulence simulations are higher than their equipartition values in the hot medium (c.f. \S\ref{sec:cold_B_field}). As discussed earlier, this higher value in the cold and mixed gas can be due to higher equipartition values in denser gas (as $B_{\rm eq} \propto \sqrt{\rho}$, \textchange{with the caveat of} assuming similar turbulent velocities in hot and cold medium, \textchange{which we discuss later in the section}), due to flux-freezing during compression from hot to cold medium \citep[e.g.][]{Sharma2010ThermalClusters, Gronke2020HowWind}, or due to magnetic draping around the cold gas clumps \citep{Dursi2008DrapingEffects,McCourt2015MagnetizedWind}.

It is hard to disentangle these three processes as the extent of amplification in the simulation ($\approx 6 B_0$) can be achieved via all above the processes. The flux-freezing can cause an amplification up to $\chi^{2/3}B_0 \approx 22 B_0$, assuming an isotropic, isobaric collapse from $T_{\rm hot}$ to $T_{\rm cold}$ and conservation of magnetic flux. The local dynamo and magnetic draping can account for an amplification up to $\chi^{1/2}B_0 \approx 10 B_0$, assuming the amplification continues until equipartition is reached, i.e. $\Ma\sim1$, and similar $v_{\rm turb}$ surrounding the cold gas means the new equilibrium magnetic field in the cold gas increases by $\sqrt{\rho_{\rm cold}/\rho_{\rm hot}}$. The flux-freezing causes a higher magnetic field in newly formed cold or mixed gas, and the other two processes amplify the existing magnetic field in the cold or mixed gas. As the magnetic fields reach equipartition values, they start to become stiff to the gas motions and start to back-react and influence the gas motions. This means the amplification value of $\chi^{1/2} B_0 \approx 10B_0$ at equipartition gives a rough upper limit on the amplification by all the processes. And, this agrees with our results in Fig.~\ref{fig:Bfield_hist}.

Out of the possible processes, turbulent local dynamo and magnetic draping are less likely due to a few reasons. For the magnetic fields to be amplified to $10B_0$ due to turbulent local dynamo, the turbulent velocity at cold gas cloud scales has to be similar to the hot gas turbulent velocity. But, due to the small scales of the cold gas clumps, the turbulent velocities at cloud scales will be much lower at $\sim v_{\rm turb}(l_{\rm clump}/L_{\rm box})^{2/3}$. Hence, the local dynamo will not be able to cause the calculated high amplifications.

For magnetic draping to amplify the fields, there needs to be a significant relative velocity ($v_{\rm rel}$) between the hot and cold gas, which generally is not the case, as we find a very similar VSF for hot and cold gas and low shear between the phases. This means, the $v_{\rm rel} \ll v_{\rm turb}$, hence the amplification of magnetic fields due to such process is probably insignificant. In addition, draping generally requires and leads to structured magnetic fields as they `drape' around the clouds \citep{Dursi2008DrapingEffects} -- something we do not observe in our simulations.

This leaves flux-freezing and subsequent compression of magnetic fields as the only process that can cause significant amplification. Once, the amplification reaches a limit where the magnetic fields are stiff (trans/sub-Alfvénic), the gas continues to evolve along magnetic field lines, hence cold gas growth does not necessarily have to compress the magnetic fields.\\

Next, we consider the structure of the magnetic fields. 
In our study, we find that the extent of entanglement ($l_{\rm stream}/l$) of the magnetic field lines increases linearly with the length of the streamline (c.f. Fig.~\ref{fig:Bfield_tangle}). This points to a structure where the longer the streamlines are, the more relatively small-scale structures are sampled. This is possible if the magnetic streamlines have a ``fractal-like'' structure that goes on until a fixed small-scale, which is the grid-scale in our simulations. Hence, the longer the streamlines are, the wider the range of perturbations that are included, leading to the increasing trend of entanglement. 

This is analogous to the well-known problem of measuring a coastline, where the measured coastline length increases with decreasing length of the measuring stick. In this case, the roles are reversed. The measuring stick has a constant length, while we make the coastline longer. Assuming self-similarity, if we rescale this longer coastline, we effectively make the measuring stick smaller and we get back to the original coastline measuring problem. Let $\epsilon \propto 1/l_{\rm stream}$ be the effective length of the measuring stick. Hence, $L = l_{\rm stream}/l$ will be the rescaled coastline (streamline) length. We can use the relation found in Fig.~\ref{fig:Bfield_tangle} and the expression for the length of self-similar fractals, i.e. $L \propto \epsilon^{1-D}$ \citep{Mandelbrot1967HowDimension, Mandelbrot1983TheNature} to find the fractal dimension ($D$) of the magnetic field lines. We find that the magnetic field lines have a fractal dimension, $D = 2$. Such naturally occurring fractal structures with a fractal dimension of 2 in 3D space are known to exist, with Brownian motion being one example \citep{Falconer1985TheSets}. Previous studies of TRMLs have also found fractal structures, for example, \citet{Fielding2020MultiphaseLayers} show that the cooling layer in a TRML has a fractal dimension of 2.5, while \citet{Tan2021RadiativeCombustion} find a slightly different value but note that measured values can differ.

We also find that a $\Gamma$ distribution on logarithmic entanglement ($\log_{10}{l_{\rm stream}/l}$) matches fairly well with the computed distribution from the simulations for longer streamlines. The $\Gamma$ distribution does a poorer job for very short streamlines, which might hint towards a transition to or altogether a different distribution for the entanglement. Or, this might possibly be due to higher resolution effects on the shorter streamlines. 

\textchange{We hope that this analytic form for magnetic field entanglement will be helpful in development of models for transport charged particles through magnetised multiphase turbulence.}







\subsection{Connection to observations}
\label{sec:disc_mock_spec}

Multiwavelength studies now allow the joined observational study of multiphase astrophysical media. Of the many ways to probe the properties of the multiphase gas, the absorption lines are one of the widely used methods \citep[e.g.,][]{2010ApJ...717..289S,Crighton2015Metal-enrichedGalaxy,2017ASSL..434..291C,2022ApJ...936..171R}. The different phases in the CGM of an intervening galaxy can deposit absorption features on the background quasar continuum. As the different sections of the absorbing medium can be moving with different velocities, these absorption features can be deposited at different Doppler-shifted positions near the line centre with different widths. Hence, the absorption features provide information about the kinematics and structure of the absorbing medium.

We find that there is no significant difference in mock absorption features of MgII 2796\r{A} with and without magnetic fields. We, furthermore, show mock absorption spectra from both HD and MHD simulations agree with observed MgII absorption features from \citet{2003AJ....125...98C,Churchill2020MgModels}, who established a relation between the number of `absorbers' and the total equivalent width of the absorption. We also show in Appendix~\ref{sec:app_mock_spectra} that this agreement is approximately valid across spectral resolution and absorption lines.

As we found a universal clump mass distribution following $\dd N/\dd m \propto m^{-2}$ (cf. Fig.~\ref{fig:cold_clump_hist} and \S~\ref{sec:cold_morph}) in both the HD and MHD cases (consistent with \citealp{Gronke2022SurvivalMedium}), this suggests that the \citet{Churchill2020MgModels} is a direct consequence of the clump mass distribution, and similar probes might be used to constrain it providing an interesting avenue for future work.

In addition to absorption lines, there are many studies that investigate the emission lines from multiphase media. \citet{Li2022TurbulenceGalaxy} look at the multiphase turbulence in the ram-pressure stripped tail of ESO 137-001 using different emission lines. They find a similar velocity structure function as ours (in Fig.~\ref{fig:vsf}) and many other simulations \citep{Mohapatra2021TurbulentMedium, Mohapatra2022VelocityClusters}. This shows that both simulations and observations point towards a high extent of kinematic coupling between the different phases in astrophysical media.


\subsection{Connection to previous studies}
Due to the very high Reynolds number of astrophysical media, they are highly susceptible to turbulence. Hence, these media are expected to be turbulent in all the different scenarios in which energy is being injected into the medium, be it via supernovae, accretion or mergers. This turbulent nature of astrophysical medium has been studied before in previous studies \citep{SchekochihinTurbulencePlasmas, Lancaster2021EfficientlyClouds, Hu2022SignatureFilaments, Li2022TurbulenceGalaxy, Federrath2013OnTurbulence,  Elmegreen2004InterstellarProcesses, Wittor2020DissectingFeedback}. There is also a plethora of studies that look at the different aspects of magnetohydrodynamic (MHD) turbulence, both in contexts related and unrelated to astrophysical mediums \citep[see, e.g., review by][]{2020arXiv201000699S}. 

Recently, there has been a significant focus on the multiphase nature of such turbulence, with or without magnetic fields. Previous studies like \citet{Mohapatra2022MultiphaseDriving} and \citet{Gronke2022SurvivalMedium} have looked into hydrodynamic multiphase turbulence, while studies like \citet{Mohapatra2022CharacterizingSimulations} and \citet{Mohapatra2022VelocityClusters} investigate the same with magnetic fields. \textchange{And, studies like \citet{2022MNRAS.514..957S} have looked at the evolution of magnetic fields in a multiphase medium}. The key difference between these studies (except \citet{Gronke2022SurvivalMedium}) and ours is the thermal instability of the ambient hot medium. In our setup, we mimic a heating source and turn off the cooling for gas hotter than $0.5T_{\rm amb}$, hence the ambient hot medium is thermally stable. Due to the absence of a thermally unstable ambient medium, mixing is the primary mechanism for creating the thermally unstable intermediate gas in our simulations. Still, results from our study will be relevant for the late evolution of simulations with thermally unstable hot medium, at which point, the further creation of cold gas is likely dominated by the cooling of mixed intermediate gas, rather than the less unstable hot medium. Importantly, the dynamics of a multiphase medium are quite different depending on which phase dominates the simulation domain. Since in most astrophysical media, the hot component is dominated by volume \citep[see, e.g.,][for the CGM]{Tumlinson2017TheMedium}, we choose to focus on the initial phase where this is also the case in our setup. Studying the full dynamic range, i.e., having a sufficiently large volume to sustain $f_{\rm V,c}\ll 1$ for an extended period of time while resolving the small-scale structure is unfortunately computationally prohibited.

Another similar system of turbulent boxes can be the stratified turbulent boxes, as studied by \citet{Mohapatra2021TurbulentMedium}, \citet{Mohapatra2021TurbulentMedium} and \citet{Wang2023TurbulentMedium}. In such systems, the fundamental nature of turbulence can be different, depending on the extent of stratification. But, due to the presence of a similar hierarchy of structure and scales, we expect to see a similar growth or destruction of cold gas. In a stratified medium, there are two kinds of motions, one across the stratified layers, i.e. along the stratifying force ($F_{\rm strat}$), and the other along the layers, i.e. perpendicular to the $F_{\rm strat}$. The growth of cold gas within the layer itself would depend on the turbulent property in the layer, roughly perpendicular to $F_{\rm strat}$, while the transport and growth of cold gas among the stratified layers would depend on the gas motion along $F_{\rm strat}$. This kind of motion can be turbulent or buoyancy-driven where the cold gas falls ``down''. A stronger stratification can suppress the turbulent motions across the stratified layers, while the buoyant forces and motions can get amplified. Hence, even though some of our results are relevant to a stratified system, due to the complex interplay between these different flows, further study is needed to fully understand the rich physics in play. 

Apart from explicitly turbulent boxes, turbulence shows up time and again in a lot of astrophysical simulations. An example of one such system are the `cloud-crushing' simulations modelling cold gas-wind interactions. These set of simulations, designed to study multiphase galactic outflows, have been extensively studied \citep[e.g.,][]{Klein1994OnClouds, Marinacci2010TheGalaxies, Scannapieco2015THECOOLING, McCourt2015MagnetizedWind, Schneider2017HYDRODYNAMICALWINDS,2021MNRAS.505.1083G}.
Studies find that cold gas clouds that are bigger than a certain critical radius can not only survive against a fast-moving hot wind but even grow as they are being entrained in the wind \citep{Gronke2018TheWind, Li2020SimulationBackground} with the details of the critical radius still under debate \citep{Kanjilal2021GrowthCooling, Farber2021TheWinds, Abruzzo2022TamingInteractions}.

Initially, when hit with the hot wind, Kelvin-Helmholtz (KH) rolls are formed near the edges facing perpendicular to the wind, where the relative velocity is the highest. These KH rolls act as one of the initial sources of turbulent motions behind the cloud in its tail and cause mixing. As the cloud gets entrained and the shear decreases, this mechanism is unable to drive any  further turbulence. Still, many of the previous studies mentioned above find that the cold gas mass continues to grow even after the cloud is entrained. This points to the presence of a substitute process for driving the turbulence at later times. The nature of this substitute process is still an open question, with some suggestions being the hot gas inflow due to cooling tail \citep{Abruzzo2022TamingInteractions} or the pulsations of the cold clumps themselves (\citet{Gronke2023CoolingCoagulation}, \citet{Gronke2020IsMisty}). Regardless of the exact source of the late-time turbulence driving in the tails, as we show in this study, if the resulting turbulence in the tail is similar, the mixing and the cold gas evolution will be similar. Interestingly, studies with magnetic fields, like \citet{Gronke2020HowWind} and \citet{Hidalgo-Pineda2023BetterDraping} find a lack of significant difference between the growth rates of the cold gas with (MHD) and without (HD) magnetic fields. This result, in combination with what we find in our study, means that the presence of magnetic fields is not affecting the turbulence-driving mechanism. However, note that \citet{Hidalgo-Pineda2023BetterDraping} do find a significant difference in the survival criterion of clouds in a laminar flow with the inclusion of magnetic fields ($\sim 2$ orders of magnitude with $\beta\sim 1$). To understand this, it is important to recall that the main difference to our turbulent setup is that for a wind tunnel setup the reduction of the drag time ($t_{\rm drag}\sim \chi v/r_{\rm cl}\sim \chi^{1/2}t_{\rm cc}$) in order to be comparable to the destruction time $t_{\rm cc}$ is sufficient for survival. \citet{Hidalgo-Pineda2023BetterDraping} attribute this reduction to a combination of draping \citep{Dursi2008DrapingEffects,McCourt2015MagnetizedWind} and an altered $\chi$ due to compression of magnetic fields. On the other hand, in a turbulent setup, the cold gas is never fully entrained.

Another analogous set of systems is the Ram-pressure stripped galaxies, also called jellyfish galaxies. Similar to the cloud-crushing simulations, such galaxies have a multiphase tail. And,  both simulations \citep{2006MNRAS.369..567R,Tonnesen2009GASMEDIUM} and observations \citep{2022A&ARv..30....3B,Li2022TurbulenceGalaxy,2023MNRAS.521.6266L} have shown the presence of turbulence in the tails of such galaxies. Results from this study will be quite relevant to the environment in such a tail, where the extent of the turbulence in the tail will dictate the overall evolution of the multiphase gas. Even though there are some strong parallels between jellyfish galaxies and cloud-crushing simulations, there are also many differences, like the difference in overdensity, presence of self-gravity, star-formation, feedback, etc. Hence, more detailed studies are required to fully understand these systems.\\

One of the major sources of turbulence in the circumgalactic medium (CGM) is the galactic outflows caused by the supernova feedback in the galactic disk. In our simulations, we vary the turbulent energy injection rate in order to get a similar turbulent velocity in both HD and MHD simulations. In a more realistic system, as in isolated galaxy simulations, the energy injection is dictated by the supernova rate, and indirectly by the star formation rate (SFR). \textchange{Previous studies like \citet{10.1093_mnras_stz3321, vandeVoort2021}} found that changes to SFR, stellar mass and ISM mass due to the inclusion of magnetic fields are small. This means the energy injection rate into the CGM is roughly unaltered due to the inclusion of magnetic fields. As the magnetic fields in the CGM will act as an additional energy sink, the resulting turbulent velocity in the CGM due to the outflows is expected to be lower when magnetic fields are included. This reduced turbulent velocity in the CGM can be one of the possible reasons for the lower extent of mixing of metals in CGM, resulting in the stronger angular dependence of metallicity in simulations when the magnetic fields are included \citep{vandeVoort2021}.

Closer to home, multiphase MHD turbulence is also seen in the solar atmosphere. The nature of MHD turbulence in the solar atmosphere is quite different, due to the very high magnetic field intensities, leading to sub-Alfvénic turbulence. In this case, the magnetic field tension is very high, and magnetic field lines are stiff to the gas flows. Still, as the mixing of multiphase gas is fundamentally tied only to the gas flows, and in the presence of the turbulent cascade of structures, our results suggest that the evolution of the multiphase gas would primarily be affected by the overall turbulent property. One of the sources for this turbulence can be the non-linear evolution of KH instability, which has been investigated in previous studies like \citet{Hillier2023TheAtmosphere}.


\subsection{Caveats / future directions}
Below, we mention some caveats of the study and some directions that can be explored in future studies.
\begin{itemize}
    \item \textit{Resolution}: We use a lower resolution in our TRML simulations compared to that in \citet{Tan2021RadiativeCombustion}. This should not affect our results because, as \citet{Tan2021RadiativeCombustion} show, it is enough to properly resolve the largest eddy to get converged cooling and mixing rates, which we do. Similarly, \citet{Gronke2022SurvivalMedium} show that the growth rates and survival of cold gas clouds well within the survival regime, is converged if the cloud radius is well-resolved. As this criterion is satisfied in our simulations, we believe the results should be converged over similar resolutions. A lack of physical resistivity, viscosity or conduction means that in our simulations these are replaced by numerical resistivity, viscosity and conduction. A higher resolution will lead to a decrease in these but, as mentioned in section~\ref{sec:disc_mass_transfer}, the primary timescale in the problem is the turbulent eddy timescale of the largest eddy, which is unaffected by the resolution. This is similar to the analogous result in TRMLs which \citet{Tan2021RadiativeCombustion} find in their study.
    \item \textit{Turbulent driving}: In this study, we maintain a solenoidal to compressive driving ratio ($f_{\rm shear}$) of 0.3 across all turbulent box simulations. Previous studies find that different $f_{\rm shear}$ in simulations can cause differences in the turbulent power spectrum \citep{Federrath2013OnTurbulence, Grete2018AsSimulations, Mohapatra2022MultiphaseDriving}. But for our results, it is enough that the turbulent eddy timescale of the largest eddy is longer than that of smaller eddies, this remains unchanged with a different turbulent driving. The nature of turbulent driving can also affect the magnetic field amplification in MHD turbulence. The magnetic field in a turbulent box driven by an $f_{\rm shear} > 0$ is amplified much faster than a purely compressively-driven ($f_{\rm shear} = 0$) turbulent box. Still, this difference is well within an order of magnitude, and the results from our simulations should largely be applicable to the case of purely compressible turbulence.
    \item \textit{Subsonic vs supersonic}: In this study, we restrict ourselves to the subsonic regime in both TRML and turbulent box simulations (to be more applicable to most astrophysical systems). \citet{Yang2023RadiativeNumbers} have looked at the behaviour of TRMLs with supersonic shear velocities, and find that for very high Mach numbers, the turbulent velocities in the mixing zone start to saturate with increasing shear velocities. This leads to a stagnation in the cooling rate, which is in agreement with our results from TRML simulations. \citet{Mohapatra2022CharacterizingSimulations}, in their simulations with supersonic turbulence, find that stronger turbulence can lead to higher compression and rarefactions in the medium. The stronger compression, along with shocks, might cause higher cold gas formation from the cooling of the ambient medium if the cooling is stronger than shock heating. This might also be valid for the supersonic multiphase turbulent boxes with non-cooling ambient medium, analogous to this study, where shocks passing through the medium might result in more efficient cooling of shocked intermediate gas regions. On the other hand, shocks in supersonic turbulence can also lead to the destruction of the cold gas, countering the additional cold gas formation. We see a hint of this more efficient destruction in our transonic ($\Ms\approx 0.9$) turbulent simulations in Fig.~\ref{fig:cloud_survival}, where the clouds larger than the subsonic critical radius get destroyed. Hence, the results in an analogous multiphase supersonic turbulent box might vary from the subsonic cases. This is further complicated by the presence of magnetic fields, where there are two kinds of shocks, and these can also lead to the amplification of the magnetic fields.
    \item \textit{Super-Alfvénic vs Sub-Alfvénic}: Most of the large-scale astrophysical media like the ISM, CGM and ICM are usually super-Alfvénic (\Ma>1) in nature. Even though, most of them start with a relatively high \Ma, due to amplification of the magnetic fields the media reach a lower \Ma, but usually not equipartition due to temporal evolution, and stay Super-Alfvénic. Similarly, in our simulations, we start well within the Super-Alfvénic regime but during the turbulent driving, we reach equipartition, before we introduce the cold gas cloud. That is, \Ma$\gtrsim$ which is tran-Alfvénic to mildly super-Alfvénic. This setup works well to understand the above mentioned astrophysical media, but there are other multiphase environments like the Solar Corona where the medium is well within the sub-Alfvénic regime and our turbulent boxes may not be analogous anymore. On the other hand, our TRML simulations include simulations with trans-Alfvénic to mildly sub-Alfvénic motions. We find that our conclusion about the relation between turbulent velocity and mixing still holds. This means, given there are turbulent motions and a turbulent cascade, the mixing will only depend on the turbulent properties and not the presence or absence of the magnetic fields. Still, we have not explicitly tested this in a turbulent box setup but can be a topic for future investigations.
    \item \textit{Anisotropic conduction}: It is well-known that conduction is anisotropic in the presence of magnetic fields. But, as we do not have physical conduction in our simulations, the numerical conduction in the simulations is isotropic in both HD and MHD cases. While it has been shown by \citet{Tan2021RadiativeCombustion,Tan2021ALayers} that generally turbulent diffusion dominates over the laminar one (thus, explaining seemingly `puzzling' convergence of larger scale studies such as ours), \textchange{this has only recently been investigated with anisotropic conduction in an MHD setup by \citet{2023arXiv230712355Z} who corroborate our results and find similar trend in suppression of cooling \citep[see, however, ][who included anisotropic conduction in their `cloud-crushing' simulations and find similar mass growth rates as the pure hydro runs]{Bruggen2023,2023MNRAS.518.5215J}}. 
    This will add an additional layer of complexity, and can also be a future direction to explore.
    \item Other effects neglected in this study are \textit{cosmic rays}, \textit{viscosity}, and geometrical variations such as \textit{stratification}. Our goal here was to study mixing in MHD in a simplified setup to which we will add additional layers of complexity in future work.
\end{itemize}


\section{Conclusions}
\label{sec:conc}

In this study, we investigate the influence of magnetic fields on the general phenomenon of mixing between the phases in a multiphase gas. For that purpose, we use two sets of simulations, turbulent radiative mixing layers (TRMLs) and turbulent boxes, with and without magnetic fields. First, we expand the parameter space for TRMLs explored in previous studies, to confirm the suppression of mixing for different cooling strengths (and hence different Damk\"ohler numbers) at different initial magnetic field orientations. Second, we check for the effects of including magnetic fields in turbulent box simulations similar to \citet{Gronke2022SurvivalMedium}. We investigate for any differences in cold gas growth rates and survival. We also study the effects of magnetic fields on the morphology of the multiphase gas and magnetic fields and check for the subsequent observational consequences.

The following are the main conclusions from this study:
\begin{itemize}
    \item We find that magnetic fields, in general, suppress the mixing in turbulent radiative mixing layers. The exception being some cases with magnetic fields are perpendicular to both shear and interface normal. This suppression is due to either amplification or the existence of strong magnetic fields along the shear, which stabilises the mixing layer.
    \item The inclusion of magnetic fields in TRML simulations only affects the generation of turbulence. We find that the relation between turbulent velocity in the mixing layer and mixing (hence cooling) rates from hydrodynamic simulations \citep{Tan2021RadiativeCombustion} still holds.
    \item We find that turbulent box simulations do not show significant differences in growth rates between identical cases with and without magnetic fields. Similarly, the survival criterion of cold gas is also unaffected by the inclusion of magnetic fields.
    \item We show that this lack of difference, with and without magnetic fields, is in line with our results from TRML simulations where the relation between the turbulence and mixing is unaffected by the presence of magnetic fields. Given similar turbulent properties, we find that the mixing between phases in a multiphase medium will also be similar, regardless of the details of turbulence generation including the presence or absence of magnetic fields.
    \item We verify that the turbulent boxes with and without magnetic fields show similar clump size distribution ($N(>m)\propto m^{-1}$), which is in agreement with previous studies. But, we find that exact morphologies are different, with the clumps being more filamentary when magnetic fields are included.
    \item We find the cold phase to be generally well entrained with the hot phase with the MHD simulation reaching this entrained state faster than the HD one. This implies that `shear-driven' mass transfer is not sufficient to explain the growth rates observed.
    \item We use mock absorption line observations of MgII to check the observational consequences of such differences in the morphology. While we do not find a significant difference between the statistics of the two cases with and without magnetic fields, both cases roughly agree with observations.
    \item We investigate the magnetic field structure in turbulent boxes. The cold gas phase has a higher mean magnetic field due to flux-freezing. We use the magnetic field streamlines to show the fractal nature of magnetic field lines and find an approximate distribution for the extent of magnetic field entanglement.
\end{itemize}

Our study reconciles the seemingly contradictory results of the effect of magnetic fields in turbulent mixing layers and a fully multiphase turbulent setup. This result also implies that the presence of cold gas in multiphase media can be explained through continuous mixing and cooling --  and this channel is not hindered by the presence of magnetic fields. However, the topic of multiphase MHD turbulence still remains full of many unanswered questions, like the effect of cosmic rays, thermal conduction, viscosity, etc. which we hope to tackle in future work.

\section*{Acknowledgements}
\textchange{The authors thank the referee for useful comments.} We thank \textchange{Amit Seta}, Brent Tan, Chad Bustard, Drummond Fielding, Eugene Churazov, Evan Schneider, Peng Oh, Ruediger Pakmor, Ryan Farber, Seok-Jun Chang, Silvia Almada Monter, Volker Springel, \textchange{Yohan Dubois and Xihui Zhao} for helpful discussions. HD thanks staff and colleagues at Max Planck Institute for Astrophysics and The International Max Planck Research School on Astrophysics for their valuable support during the research. 
 MG thanks the Max Planck Society for support through the Max Planck Research Group.
 Computations were performed on the HPC system Freya and Cobra at the Max Planck Computing and Data Facility.

This research made use of \texttt{Athena++} \citep{Stone2009AMHD}, \texttt{yt} \citep{Turk2011Yt:DATA}, \texttt{Trident} \citep{Hummels2017Trident:Simulations}, \texttt{NumPy} \citep{Harris2020ArrayNumPy}, \texttt{matplotlib} \citep{Hunter2007Matplotlib:Environment}, \texttt{SciPy} \citep{Virtanen2020SciPyPython} and \texttt{NetworkX} \citep{Hagberg2008ExploringNetworkX}.

\section*{Data Availability}
Data related to this work will be shared on reasonable request to the corresponding author.



\bibliographystyle{mnras}
\bibliography{refs_new}

\appendix

\section{Comparison of methods to calculate \lowercase{$u^\prime$} }
\label{sec:uprime_method_appendix}

\begin{figure}
	\includegraphics[width=\columnwidth]{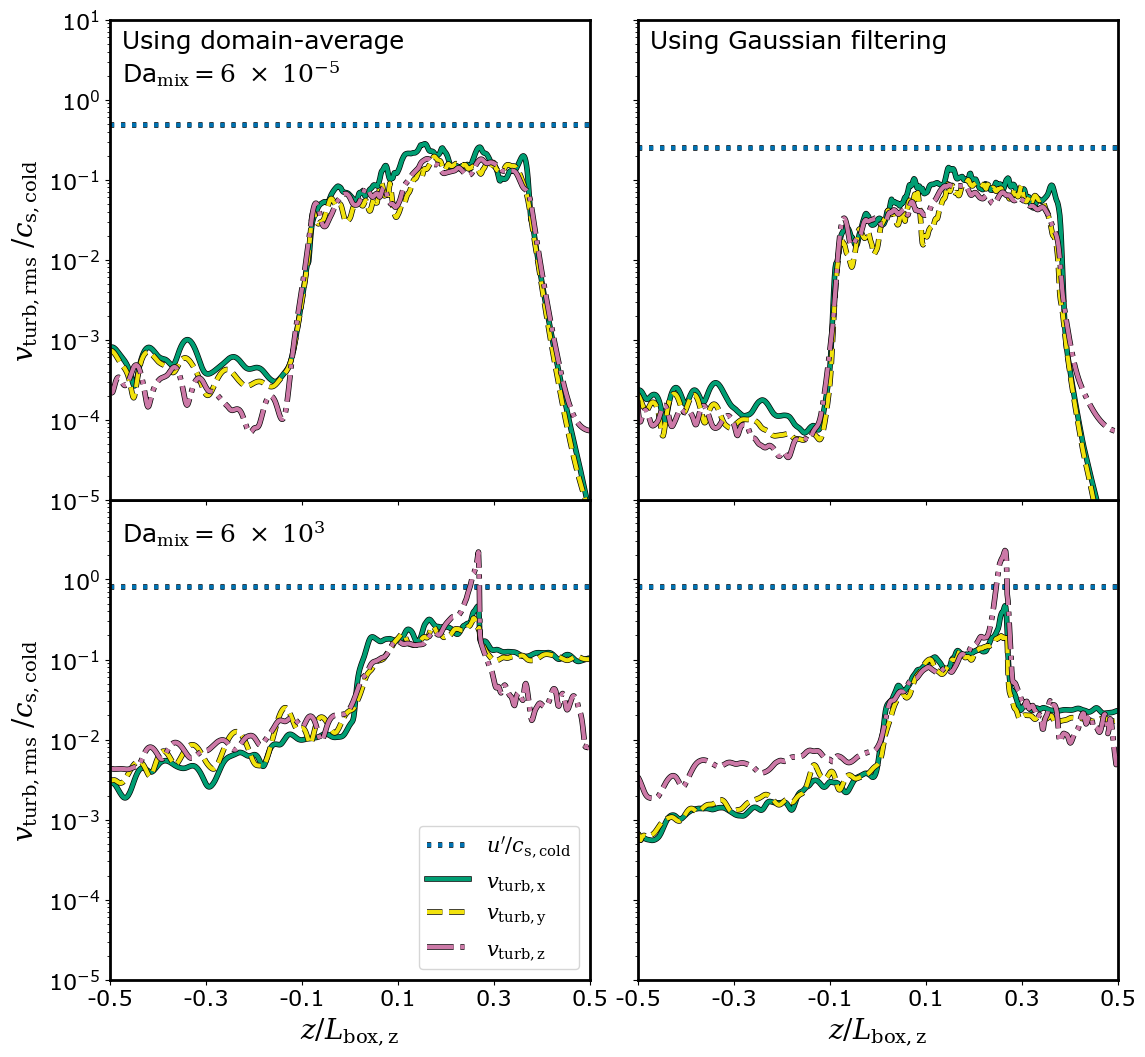}
    \caption{$u^\prime$ profiles for two different methods. The left panel shows the averaging method used in \citet{Tan2021RadiativeCombustion}, and the right panel shows the Gaussian filtering method used in \citet{Abruzzo2022TamingInteractions}. We find only minor differences between the two methods which, at worst, stay within an order of magnitude.}
    \label{fig:uprime_profiles_gauss}
\end{figure}

\begin{figure}
	\includegraphics[width=\columnwidth]{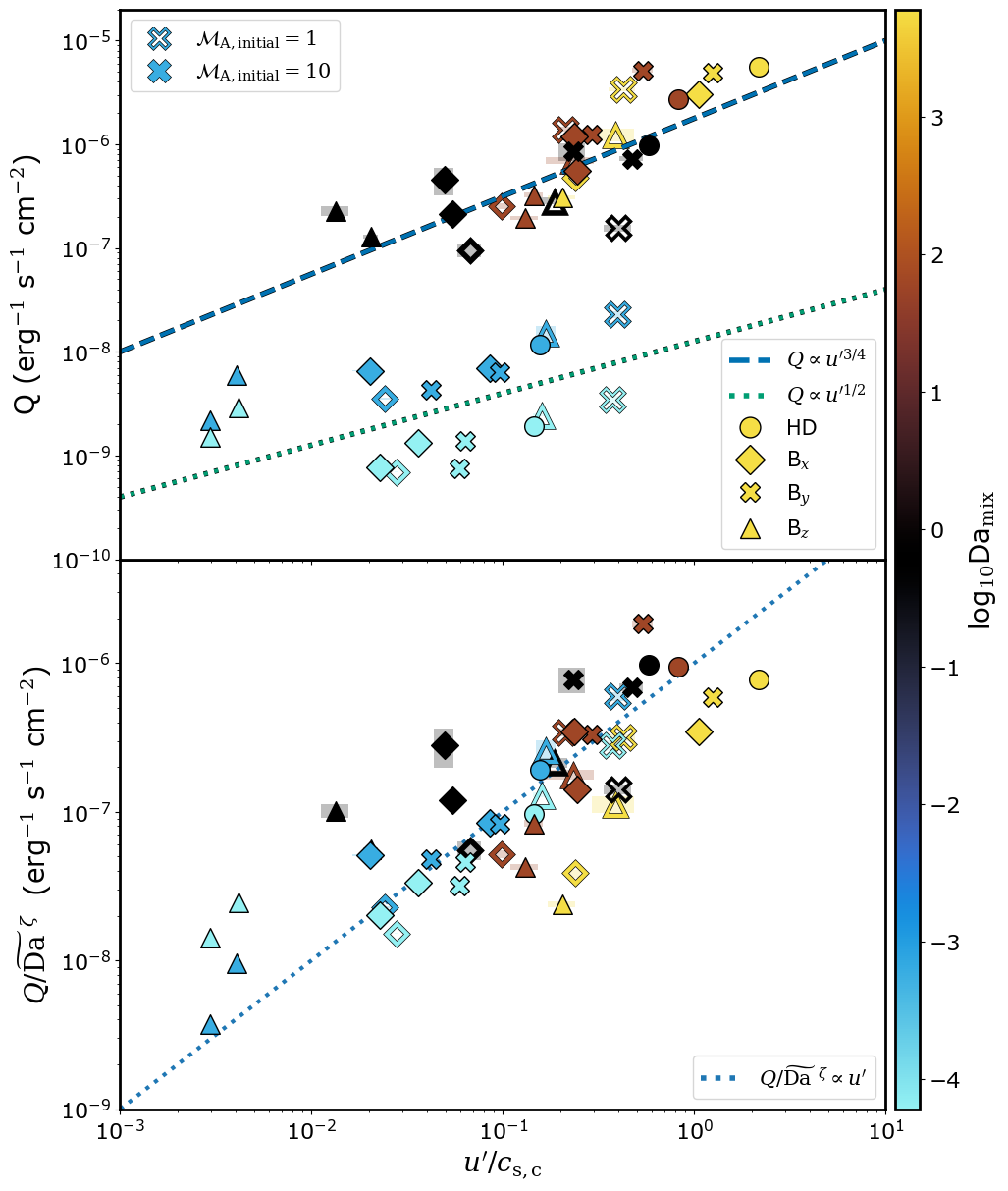}
    \caption{Same figure as Fig~\ref{fig:Q_vs_vturb}, but the $u^\prime$ is calculated using the Gaussian filtering method from \citet{Abruzzo2022TamingInteractions}. This shows that the results in Fig~\ref{fig:Q_vs_vturb} are not sensitive to the method used to calculate the turbulent velocity.}
    \label{fig:Q_vs_vturb_gauss}
\end{figure}

There are different ways to calculate the turbulent velocity ($u^\prime$) in TRML simulations, and it is important to ensure that our conclusions are not sensitive to the choice of the method. In this section, we compare two methods of calculating $u^\prime$. The first method is the one we use for the analysis in this paper and is the same method used in \citet{Tan2021RadiativeCombustion}. The second method is employing Gaussian filtering \citep[e.g.][]{1994PhFl....6.1775B,2000ExFl...29..275A,Abruzzo2022TamingInteractions}. Fig~\ref{fig:uprime_profiles_gauss} shows the same analysis as Fig~\ref{fig:uprime_profiles} but using the two methods, and Fig~\ref{fig:Q_vs_vturb_gauss} shows the same analysis as the Fig~\ref{fig:Q_vs_vturb} but using the $u^\prime$ calculated using the Gaussian filtering method.

We find that our results are robust across the two methods and are insensitive to the differences between these two methods of calculating $u^\prime$.

\section{Quantification of filamentariness}

As mentioned in \S~\ref{sec:cold_morph}, we use neighbourhood graphs for each cold gas clump to calculate the measure of filamentariness. The following are the different steps we take to calculate the measure, after we use a clump-finding method to identify the cold gas clumps:
\begin{enumerate}
    \item Calculate the adjacency matrix for each gridcell inside the clump. If a speedup is needed, construct another adjacency matrix for each $n^\text{th}$.
    \item Construct the neighbourhood graphs from all the adjacency matrices constructed in the previous step.
    \item Calculate the shortest path between each node in the smallest of the neighbourhood graphs created in the previous step.
    \item Find the longest of the set of calculated shortest paths and note the nodes corresponding to that path.
    \item Recalculate the length of the longest ``shortest'' path between the nodes from the previous step, using the largest neighbourhood graph.
    \item The length from the previous step gives a rough measure of the filament length in the clump. Repeat the steps for all clumps.
\end{enumerate}

We tested the above method for different numbers of skipped points for calculating the shortest paths. We find a negligible difference in the calculated length of large clumps even up to the point where every $20^\text{th}$ point is considered. We see major deviations only when the skipped points are a big majority of the points and the resulting neighbourhood graph is not representative of the clump anymore.

\begin{figure}
	\includegraphics[width=\columnwidth]{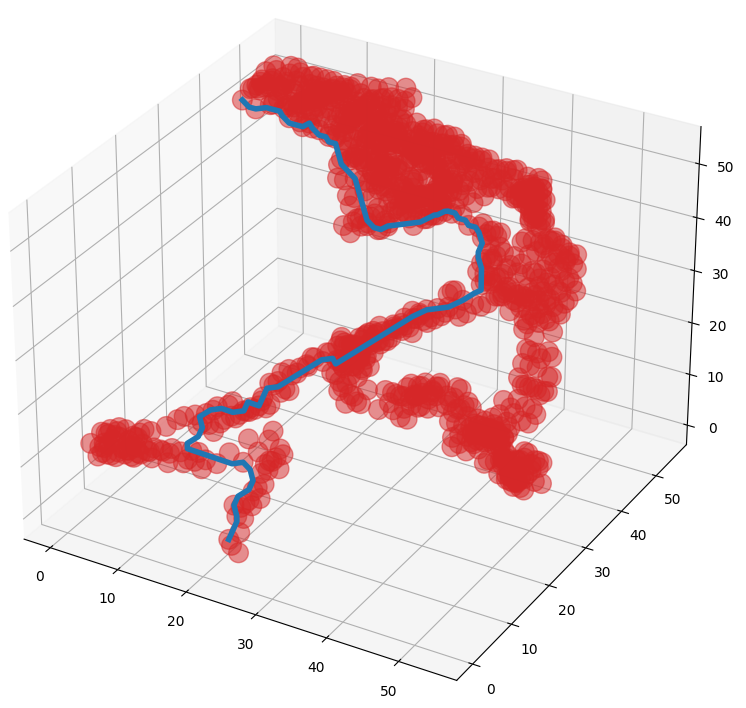}
    \caption{Red points show the points from the skipped graph of a single clump, and the blue solid line shows the calculated filament length using every $4^\text{th}$ point. Axis labels correspond to the number of gridcells.}
    \label{fig:fil_algo}
\end{figure}

In this work, we only skip every $4^\text{th}$ point in the clump. Fig.~\ref{fig:fil_algo} shows an example of the calculated filament length for a clump in an MHD turbulent box simulation.

\section{Mock spectra}
\label{sec:app_mock_spectra}

\begin{figure}
	\includegraphics[width=\columnwidth]{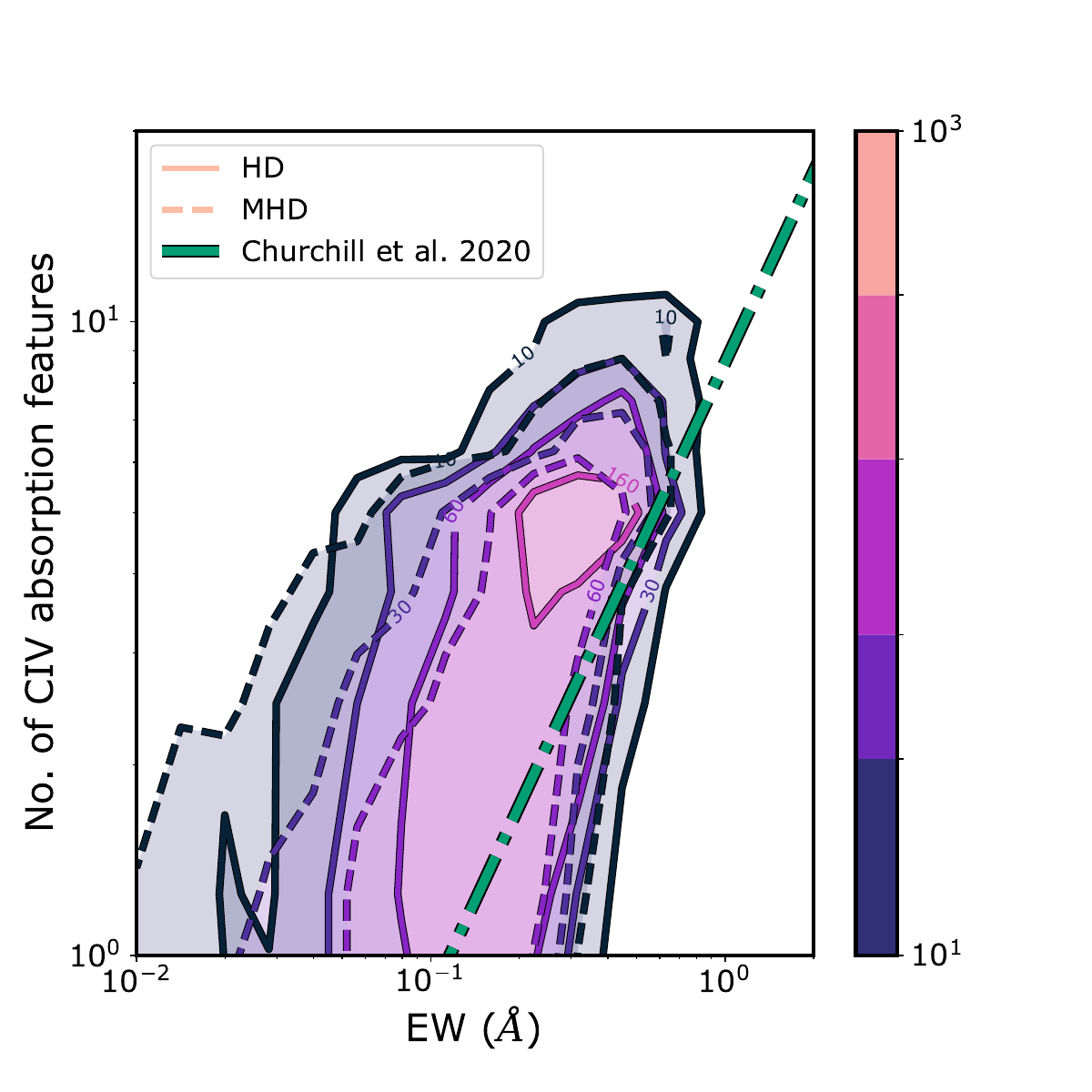}
    \caption{Same as Fig.~\ref{fig:los_churchill} but for high resolution MgII 2796 \r{A} absorption spectra with $d\lambda = 0.01$\r{A}}
    \label{fig:MgII_highres}
\end{figure}

\begin{figure}
	\includegraphics[width=\columnwidth]{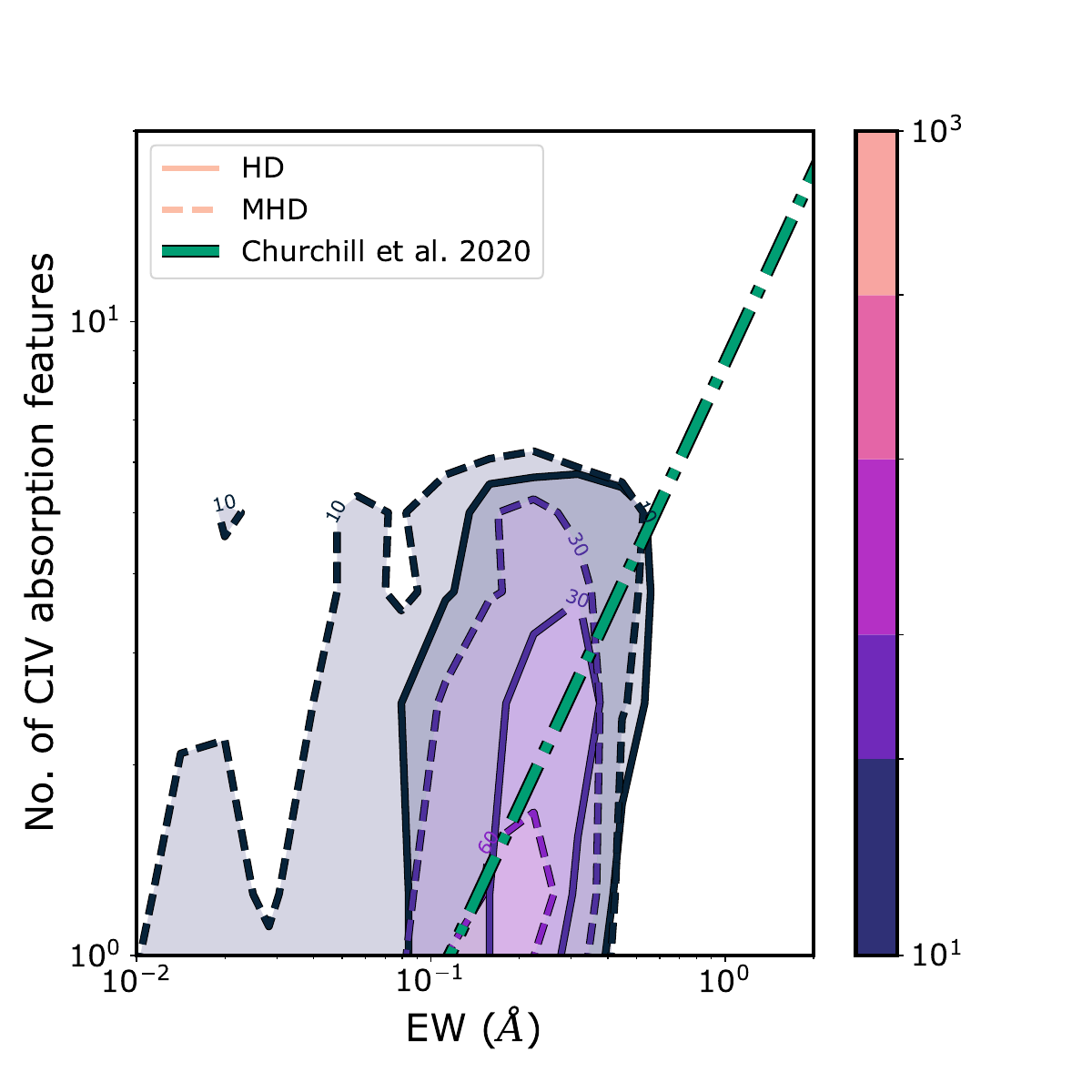}
    \caption{Same as Fig.~\ref{fig:los_churchill} but for high resolution CIV \r{A} absorption spectra with $d\lambda = 0.01$\r{A}}
    \label{fig:CIV_highres}
\end{figure}

As mentioned in \S~\ref{sec:mock_spec} and \ref{sec:disc_mock_spec}, we find only marginal differences between the statistics of the MgII mock absorption spectra of the HD and MHD simulations, despite significant differences in the column densities. This can be due to the specific property of the MgII 2796\r{A} absorption line, like the curve of growth flattening around similar column densities, which can lead to smaller differences. Another possible reason for this lack of difference can also be the spectral resolution. To address both these points, we first increase the spectral resolution of the mock spectra tenfolds to $\Delta\lambda=0.01$\r{A} and recreate the same MgII 2796\r{A} mock absorption spectra analysis as Fig.~\ref{fig:los_churchill}. Secondly, we repeat the same analysis for CIV 1551\r{A} at the higher resolution. As a significant fraction of the CIV mock absorption lines are saturated, we use a more relaxed constraint for the minimum (0.01) and maximum (0.95) absorbed flux. Fig.~\ref{fig:MgII_highres} and \ref{fig:CIV_highres} show the results from the analysis of higher spectral resolution MgII 2796\r{A} and CIV 1551\r{A} mock absorption spectra.

We find that the increase in spectral resolution of mock MgII absorption spectra shifts the relation between the number of features and total equivalent width, but it roughly follows the same slope as the observed relation from \citet{Churchill2020MgModels}. Surprisingly, the statistics of the mock CIV 1551\r{A} absorption spectra also seem to agree with the observed MgII relation, and the HD-MHD differences are wider as expected, but to the lower number of unsaturated mock spectra, it is harder to draw concrete conclusions.

This apparent robustness of the observed relation might hint towards a more fundamental origin of the relation, like the clump distribution. But, we leave it to future studies to investigate this further.

\textchange{\section{Effect of stochasticity}}
\label{sec:app_stochasticity}

The stochastic nature of the turbulence can cause variations in the evolution of quantities in a turbulent environment. \citet{Gronke2022SurvivalMedium} found that this stochasticity affects the cold gas mass evolution in hydrodynamic turbulent boxes with an intermediate-sized initial cold cloud. In this regime, they saw both survival and destruction of the cold cloud for different choices of random seeds for turbulent driving. This was attributed to the higher significance of the exact turbulent velocity field in the intermediate regime between cloud survival and destruction.

We repeat this test for our simulations, with  and without magnetic fields. We run turbulent box simulations with 3 different random seeds at $\mathcal{M} = 0.5$ and introduce clouds of different sizes to check for the effect of stochasticity of the turbulence. We use a $L_{\rm box}/R_{\rm cl} = 20$, instead of 40, due to its lower computational costs.

\begin{figure*}
	\includegraphics[width=\textwidth]{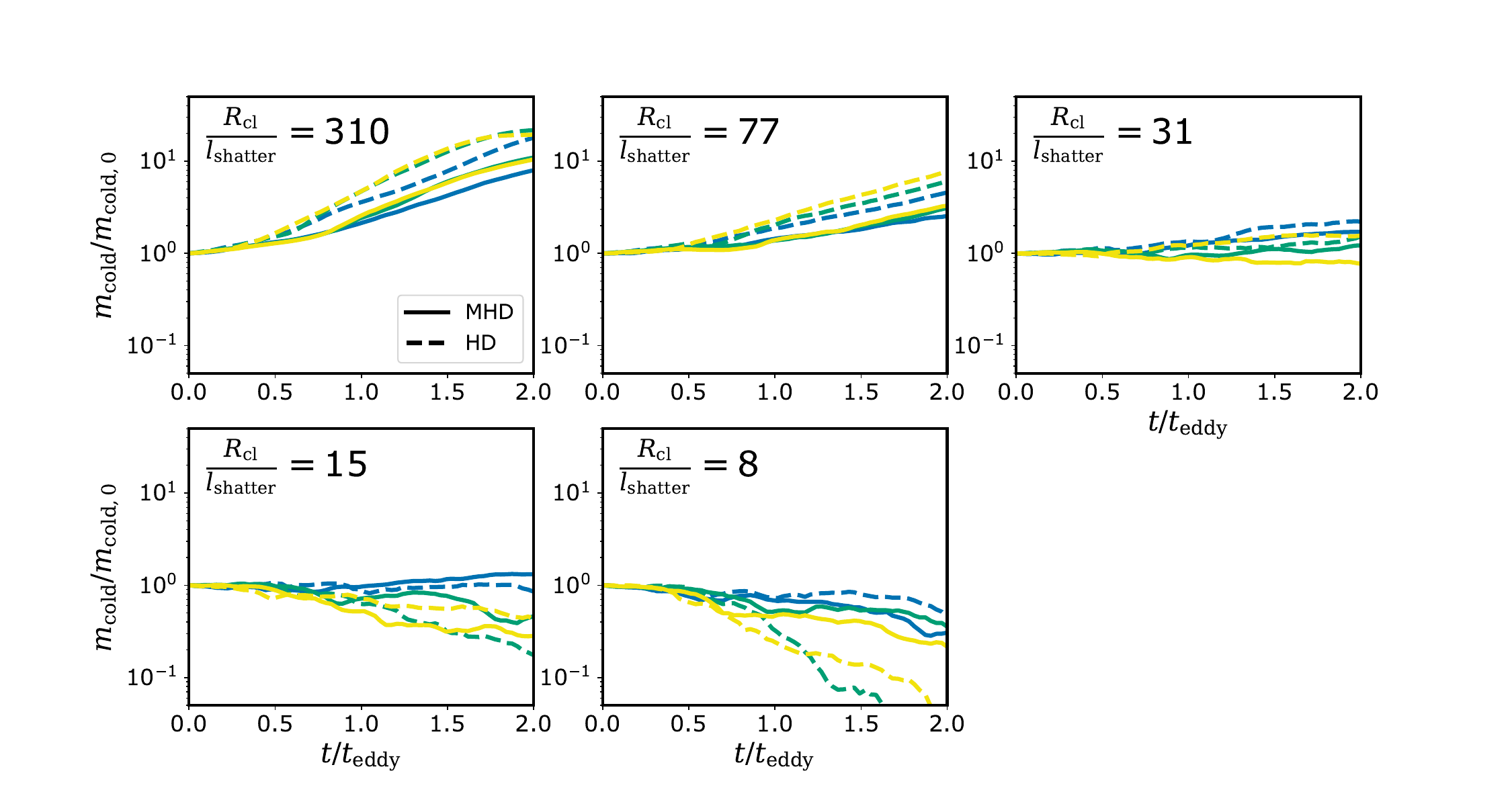}
    \caption{Cold gas mass evolution for simulations with the same parameters but different random instances of turbulence. The different panels refer to different $R_{\rm cl}/l_{\rm shatter}$ in a turbulent medium with $\mathcal{M} = 0.5$. The different colours denote simulations with varying random seeds for turbulent driving. The solid and dashed lines show the evolution of simulations with and without magnetic fields, respectively.}
    
    \label{fig:stochasticity}
\end{figure*}

Fig.~\ref{fig:stochasticity} shows the cold gas mass evolution for the different cases. We find that the cold gas mass growth/destruction rate for cold gas clouds in intermediate and destruction regimes is sensitive to the exact choice of the random seed. We also find that this is true for both HD and MHD and with no clear order of growth rate between the HD and MHD counterparts. This high dependence on stochasticity in these regimes is due to the lack of cold gas mass. This results in a very stochastic sampling of turbulence, hence making the evolution very stochastic in nature.

On the other hand, in the survival regime, the MHD simulations seem to have a slightly lower growth rate, compared to their HD counterparts, although still a much lower difference compared to the order of magnitude difference observed in TRML simulations. We attribute this minor difference to some unavoidable systematic differences between the HD and MHD simulations. The biggest of them is the difference in dissipation rate between MHD and HD, due to the extra dissipation of magnetic energy via numerical resistivity. This higher dissipation results in a slightly hotter medium in a fully developed turbulent box, in turn resulting in a slightly deviated density distribution. These slight deviations affect the evolution via a slight difference in overdensity, mixed gas temperature, etc.

Still, as Fig.~\ref{fig:stochasticity} shows, this difference is minor and it gets even more trivial when we take the spread due to stochasticity into account.

\bsp	
\label{lastpage}
\end{document}